\documentclass[preprint2]{aastex62}
\usepackage{commands_astral}
\usepackage{graphicx}
\usepackage{longtable}
\usepackage{url}
\usepackage{color}
\usepackage[utf8]{inputenc}

\shorttitle{ASTRAL M Star Reference Spectra}
\shortauthors{Carpenter et al.}

\begin{document}

\title{The Advanced Spectral Library (ASTRAL): Reference Spectra for Evolved M-Stars\footnote{Based on observations with the NASA/ESA {\it{Hubble Space Telescope}} obtained at the Space Telescope Science Institute, which is operated by the Association of Universities for Research in Astronomy, Incorporated, under NASA contract NAS5-26555.}}

\author{Kenneth G. Carpenter}
\affil{NASA/GSFC Code 667, Goddard Space Flight Center, Greenbelt, MD 20071}

\author{Krister E. Nielsen}
\affil{Catholic University of America, Washington, DC 20064}

\author{Gladys V. Kober}
\affil{Catholic University of America, Washington, DC 20064}
\affil{NASA/GSFC Code 667, Goddard Space Flight Center, Greenbelt, MD 20071}

\author{Thomas~R.~ Ayres}
\affil{University of Colorado, Boulder, CO 80309} 

\author{Glenn~M. Wahlgren}
\affil{General Dynamics Information Technology, 3700 San Martin Dr, Baltimore, MD 21218} 

\author{Gioia Rau}
\affil{NASA/GSFC Code 667, Goddard Space Flight Center, Greenbelt, MD 20071}
\affil{Catholic University of America, Washington, DC 20064}

\begin{abstract}
The \hst\ Treasury Program ``Advanced Spectral Library Project: Cool Stars"  was designed to collect representative, high quality  ultraviolet  spectra of eight evolved F$-$M type cool stars. The Space Telescope Imaging Spectrograph (STIS) echelle spectra of these objects enable investigations of a broad range of topics including stellar and interstellar astrophysics. This paper provides a guide to the spectra of the two evolved M-stars, the M2Iab supergiant \aori\ and the M3.4 giant \gcru, with comparisons to the prototypical K1.5 giant \aboo. It includes identifications of the significant atomic and molecular emission and absorption features and discusses the character of the photospheric and chromospheric continua and line spectra. The fluorescent processes responsible for a large portion of the emission line spectrum, the characteristics of the stellar winds, and the available diagnostics for hot and cool plasmas are also summarized. This analysis will facilitate the future study of the spectra, outer atmospheres, and winds, not only of these objects, but for numerous other cool, low-gravity stars for years to come.
\end{abstract}     
\keywords{stars: atmospheres -- stars: chromospheres -- stars individual (\aori, \gcru, \aboo) -- stars: late-type -- stars: mass loss -- supergiants - giants}

%\citep{Carpenter14a}
 
\section{Introduction}
The {\it Hubble Space Telescope} {\hst}  Treasury Program Advanced Spectral Library (ASTRAL) Project: Cool Stars  (PI = T. Ayres) recorded  high-resolution, high signal/noise spectra of the complete  far to near ultraviolet wavelength region of stars representing different parts of the evolution of  massive cool stars. The spectral library includes spectra for two evolved M-stars, \aori\  and \gcru, chosen as representatives  for the giant and supergiant luminosity classes. This paper provides an overview of these two reference objects with the intent to provide the tools and basic understanding of evolved stars to assist future mining of the ASTRAL spectral resource.  

The outer atmosphere and wind for the giants and supergiants are, in general, not well understood as they show significant inhomogeneities in their thermal and kinematic structure. In a step toward improved knowledge of these objects, this paper presents identifications of the significant atomic and molecular emission and absorption features observed in \aori\ and \gcru\ spectra and discusses the character of the photospheric and chromospheric continua. We present the identified fluorescence processes with initial conclusions regarding what {information they} provide on the physical conditions in the plasma.

The M2~Iab supergiant \aori\ (Betelgeuse=HD 39801) is the iconic red supergiant with very clumpy surface convection \citep{Haubois09a}, distant circumstellar shells \citep{Bernat78a}, and prominent far-ultraviolet absorption in carbon monoxide \citep{Carpenter94c}. It is an extreme object in terms of low surface temperature and gravity, high visual luminosity, and lack of coronal signatures and is an important transitional object between stars with strong corona and weak winds to objects  with weak corona and strong winds. The full ultraviolet \aori\ spectrum  was investigated using \hst/Goddard High Resolution Spectrograph (GHRS) data (R$\sim$30,000) in the near-ultraviolet \citep{Brandt95a}. Its ultraviolet spectrum is complex and has weak signatures from the photosphere but is dominated by warm chromospheric continuum and line spectrum. The emission spectrum shows signatures from fluorescence caused in particular by a strong  \lya\ \citep{Carpenter94c}. Other work on \aori\ are presented by \citet{Weymann62a, Boesgaard75a, Boesgaard79a, Goldberg84a} using ground based observations; \citet{Carpenter84a} with the \iue, \citet{Lobel00a, Carpenter94c, Carpenter97a} using \hst/GHRS data, and \citet{Lobel01a} with \hst/STIS.
 
The M3.5 III giant star \gcru\ (HD 108903)  is like \aori\ an extreme non-coronal object with similar surface temperature, but with a higher surface gravity. Gamma Cru displays a cleaner ultraviolet  spectrum  with more narrow, less blended, but still strong chromospheric and wind emission lines \citep{Carpenter95a}.  The object is a bridge-star to the warmer, non-coronal K-giants, and its clean spectrum is an important guide and tool to understanding both the spectra of the warmer giants and especially the more complex spectrum of \aori, in which the lines suffer significantly more blending with each other and mutilation by overlying absorption. Gamma Cru was observed with \iue\ (R$\sim$20,000) at most ultraviolet wavelengths \citep{Carpenter88a}, though with much lower sensitivity and less depth than is possible with \hst/STIS. Below 2200~\AA, only the strongest of the emission features, and no continuum, are visible in the \iue\ spectrum.  Higher spectral resolution and  signal-to-noise  observations with limited spectral coverage were obtained with \iue\ \citep{Carpenter88a} and GHRS (\cite{Carpenter95a}; \cite{rau18b}) and allowed for time-domain studies of the features in these limited wavelength regions. Gamma Cru's ultraviolet spectrum is dominated by chromospheric emission, but the photospheric spectrum is visible longward of  2600~\AA. The fluorescent spectrum in \gcru\ is stronger compared to \aori\ while no circumstellar shells are observed toward this giant star. 

We present the \hst\ ASTRAL  spectra and some initial results of the analysis. Further papers in this series will include studies of the  wind and associated mass-loss from these stars.

\section{Observations and Stellar Parameters}
This paper focuses on the evolved cool stars \gcru\ and \aori\ and compares their spectral characteristics to the well-studied  K1.5~III star  \aboo\ \citep{Hinkle05a}. The observational strategy for \aori\ and \gcru, as for  the other objects in the  ASTRAL Cool Star Program, was crafted to find the optimal combination of spectral resolution and signal-to-noise over the complete far- and near-ultraviolet spectra (1150$-$3159~\AA). 

\begin{deluxetable*}{lcccccc}
\tablecolumns{7}
\tabletypesize{\scriptsize}
%\rotate
\tablecaption{STIS ASTRAL Observations for \aori\  and \gcru\  \label{tbl-1}}
\tablewidth{0pt}
\tablehead{\colhead{Target} & \colhead{Grating} &\colhead{$\lambda_c$} &\colhead{Aperture} & \colhead{Spectral Coverage} & \colhead{$t_\mathrm{exp}$} & Mean Julian Date\  \\
\colhead{} & \colhead{}  & \colhead{(\AA)}  & \colhead{($''$)} & \colhead{(\AA)}  & \colhead{(s)}  & (+2,550,000)}
\startdata
\aori\  &  E140M  & 1425 &  0.20$\times$0.20 & 1150$-$1700 & 16132  &   5599.473 \\
& E230M & 1978 &  & 1640$-$2150 & 11400  & 5599.525  \\ 
&& 2707 &  & 2865$-$3118 &  1862 & 5617.910  \\
& E230H & 2263 &  0.20$\times$0.09 & 2132$-$2393 &  7554 & 5615.899 \\
&& 2513 & & 2386$-$2649 & 10404  & 5600.338 \\
&& 2762  & & 2622$-$2875 & 4704  & 5600.391  \\ [2mm]
\gcru\  &  E140M &  1425 &  0.20$\times$0.20 & 1150$-$1708 & 17751 &  5774.729 \\
  & E230M & 1978 &  & 1640$-$2150 & 12480  & 5774.663 \\ 
  && 2707 &  &  2863$-$3118 &  2131  & 5780.470 \\
 & E230H & 2263 & 0.20$\times$0.09 & 2135$-$2395 & 8363  & 5774.550  \\
 && 2513 &   & 2387$-$2647 & 8363 & 5774.596   \\
 && 2762  &  & 2623$-$2874 & 5243  & 5777.726 \\
\enddata
\tablecomments{All of these spectra consist of co-added data sets, and the Mean Julian Date  listed represents the starting time of the first segment in the observing sequence.  A more comprehensive and detailed description of these observations can be found at: http://casa.colorado.edu/$\sim$ayres/ASTRAL/. }
\end{deluxetable*}

All the spectra for the two objects are splined in a top-level dataset covering the entire wavelength region. These combined spectra are constructed so that the maximum signal-to-noise is achieved in each spectral region, for example, by co-adding spectra with high (E140H) and medium (E140M, if available) spectral resolution, after filtering the high-resolution spectrum with the low-resolution point-spread function, and vice versa. This produces a {\it hybrid}  line spread function, which does not influence the analysis of broad-line objects but is significant for narrow-line stellar and/or interstellar features. In such a case, one should examine the highest-resolution co-added spectrum. An overview of the data used in this paper is presented in Table~\ref{tbl-1} and a more detailed description of the data, including the observing strategy, data reduction, co-addition, and splicing, can be found on the ASTRAL website\footnote{http://casa.colorado.edu/$\sim$ayres/ASTRAL/} and at the Mikulski Archive for Space Telescopes (MAST)\footnote{https://archive.stsci.edu/prepds/astral/}  (\dataset[doi:10.17909/T9P016]{https://doi.org/10.17909/T9P016}). 

\citet{Lobel00a} derived $T_\mathrm{eff}$ = 3500~K and log\,$g$=$-$0.5 for \aori\  based on spectral synthesis calculations at near-infrared wavelengths.  These values are in  agreement with earlier studies by \citet{Carpenter94c} and \citet{Blackwell77a} but are slightly larger than \citet{Judge91a}. The accuracy of the basic stellar parameters for \aori\ is hampered by uncertain values for the stellar mass, radius and distance. \citet{Harper08a} suggested that \aori\ is further away than indicated by {\em Hipparcos} measurements and used the bolometric flux and angular diameter from \citet{Perrin04a} to obtain $T_\mathrm{eff}$= 3650~K and log\,$g$=$-$0.26. The derived effective temperature of \aori\ in various studies, including \citet{Martinez11a},  converge toward $\sim$3650~K,  while the surface gravity is more uncertain.

\begin{deluxetable*}{ccclcccr}
\tablecolumns{8}
\tabletypesize{\scriptsize}
\tablecaption{ASTRAL M-Star Targets\label{Table1}}
\tablewidth{0pt}
\tablehead{
\colhead{HD Number} & \colhead{Proper Name}  & 
\colhead{$T_\mathrm{eff}$(K)} & \colhead{$\log\,(g)$}  & 
\colhead{V (mag)}  & \colhead{B-V (mag)} & \colhead{Type}  & \colhead{$\Pi$~ (")} }
\startdata
 39801 & \aori\ (Betelguese) & 3650\tablenotemark{a} & $-$0.26\tablenotemark{a} & \phantom{$-$}0.42 & 1.85  & M2Iab & 0.006 \\ 
 108903 &  \gcru\ (GaCrux) &  3689\tablenotemark{b} & \phantom{$-$}0.3\tablenotemark{c} & \phantom{$-$}1.63 & 1.59  & M3.5III & 0.037 \\
 124897 &  \aboo\ (Arcturus) & 4286\tablenotemark{d} & \phantom{$-$}1.7\tablenotemark{d} & $-$0.04 & 1.23  & K1.5III & 0.089 \\
  \enddata
\tablenotetext{a}{\citet{Harper08a}}
\tablenotetext{b}{\citet{McDonald17}}
\tablenotetext{c}{\citet{Judge91a}}
\tablenotetext{d}{\citet{Ramirez11a}}
\end{deluxetable*}

Gamma Cru was also included in the study by \citet{Judge91a} and was assigned a $T_\mathrm{eff} $= 3626~K, $\log{g}$=0.3. More recently, \citet{Martinez11a} derived effective temperatures for a large number of evolved stars, including all three objects in this study, based on basal chromospheric \ion{Mg}{2} $k + h$ fluxes measured in \iue\ spectra.  The effective temperature for \gcru\  from \citet{Martinez11a} is 3577~K. \cite{Ohnaka11} presented VLTI/AMBER observations showing a photospheric effective temperature of  $T_\mathrm{eff}$= 3690$\pm54$~K, from the limb-darkened disk diameter derived from the $K$-band continuum data.

Alpha Boo (K1.5~III) is  one of the brightest stars in the northern sky. \citet{Ramirez11a} undertook a comprehensive analysis of \aboo\ based on model fits of visible to mid-IR spectrophotometric data and derived $T_\mathrm{eff}$= 4286~K. A stellar mass (1.08~$M_\sun$) and radius (25.4~$R_\sun$) for \aboo\ followed from isochrone fitting and previously reported angular diameter and distance measurements, respectively; which in turn implied a surface gravity of log\,$g$=1.66. 

%\citeauthor{Martinez11a} derived a lower value, $T_\mathrm{eff}$= 3645~K. - wrong,not for Alpha Boo! (closer to Gam Cru!)

The basic stellar parameters of the objects analyzed in this paper are presented in Table \ref{Table1}. 

\section{Turbulence}

\begin{figure}[h]
\includegraphics[width=\columnwidth]{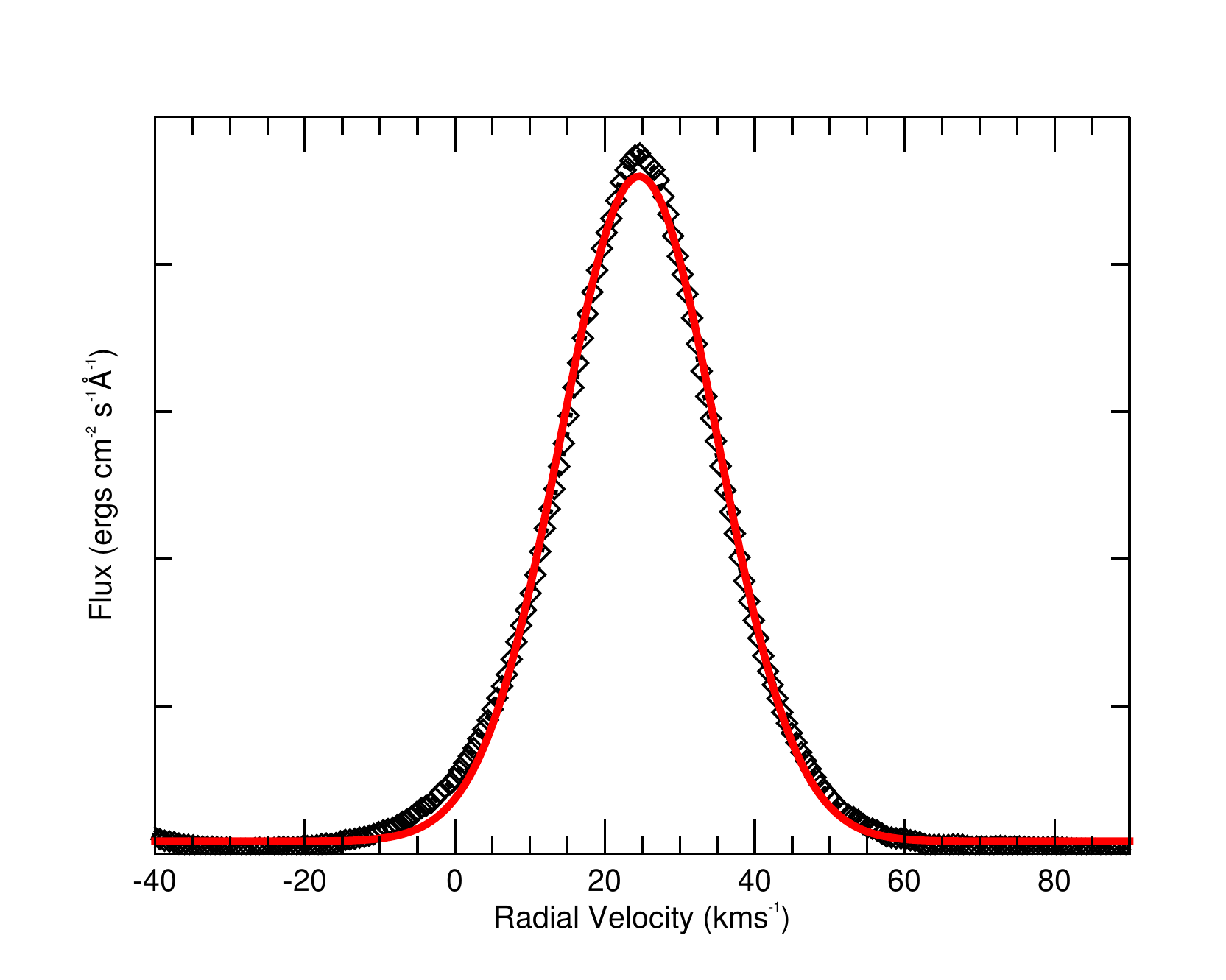} 
\caption{Gauss fit of the \ion{C}{2} \lm2325.40 in the \gcru\ spectrum. The asymmetry (enhancement) of the blue wing  that was observed in the 1995 GHRS data \citep{Carpenter95a}, and is strong in the  \aori\ spectrum, is weak in this 2011 STIS spectrum of \gcru\ (diamonds: observed spectrum, red line: fitted spectrum.) \label{CIIprof}} 
\end{figure}

{The  \ion{C}{2} intercombination lines at 2325 \AA\ have very low transition probabilities and have been  utilized in the past to derive the turbulence in non-coronal K and M stars due to their low opacity \citep[$\tau_0<$1,][]{Carpenter95a, judge86a}.  We have measured the C II lines} in the STIS ASTRAL and GHRS pre-COSTAR (i.e., before the aberration correction was installed on {\it{HST}}) \gcru\ spectra.  The FWHM is corrected for the instrumental profile, STIS 2.2 \kms, based on fitting the Line-Spread-Function (LSF) profile at 2400 \AA. The instrumental correction for the pre-costar GHRS spectrum is $\sim$15 \kms \citep{Carpenter94c}. The line asymmetry, as observed in \citet{Carpenter95a},  is weak (see Figure~\ref{CIIprof}) and may be attributed to a fluorescent \ion{Fe}{2} line, and no asymmetry is observed in the {other \ion{C}{2} lines}. The measurements {are presented in Table~\ref{turbulence}}. 

%The weak \ion{Co}{2} are omitted from the analysis due to signs of self-reversal in the \ion{Co}{2} spectrum. 

\begin{table}[h]
\caption{Turbulence measurements for \gcru}
\begin{tabular}{lcccc} \hline \hline
& \multicolumn{2}{c}{RV (\kms)} & \multicolumn{2}{c}{FWHM (\kms) }  \\ 
Line (\AA, vac.)  &  STIS   & GHRS & STIS &  GHRS \\ \hline
% \ion{C}{1} \phantom{\textsc{i}}\lm2478.56   &   19.8    &  29.0    &  22.5    &  27.6 \\
 \ion{C}{2} \lm2324.21   &   24.8    &  22.2    &  21.9    &  22.9 \\
 \ion{C}{2} \lm2325.40   &   27.7   &  25.2    &  23.8    &  27.6 \\
 \ion{C}{2} \lm2326.11   &   24.6    &  24.9    &  23.3    &  24.7 \\
 \ion{C}{2} \lm2327.65   &   25.2    &  26.5    &  23.2    &  25.5 \\
 \ion{C}{2} \lm2328.84   &   25.7    &  25.2    &  24.2    &  23.6 \\ \hline
 Mean  & 25.6 & 24.8 & 23.3 & 24.9 \\ 
 \hline
 \hline
 \end{tabular}
\label{turbulence}
\end{table}

\section{Characterization of the Spectra}
Alpha Ori and \gcru\ both show complex spectra that are dominated by a chromospheric continuum and features from the far-ultraviolet (FUV) well into the near-ultraviolet (NUV).  Beginning around 2600 \AA, the photospheric continuum becomes visible, though chromospheric emission features continue to dominate the spectra to well above 2800 \AA. The chromospheric emissions, chromospheric/photospheric absorptions, and the wind and molecular absorptions are observed  superposed over the chromospheric continuum at short wavelengths and over the photospheric continuum at long wavelengths.  In  \aori, the spectrum is made more complex by overlying circumstellar absorption.  

\begin{figure*}[h]
\begin{center}
\includegraphics[bb=26 10 703 508, width=\textwidth]{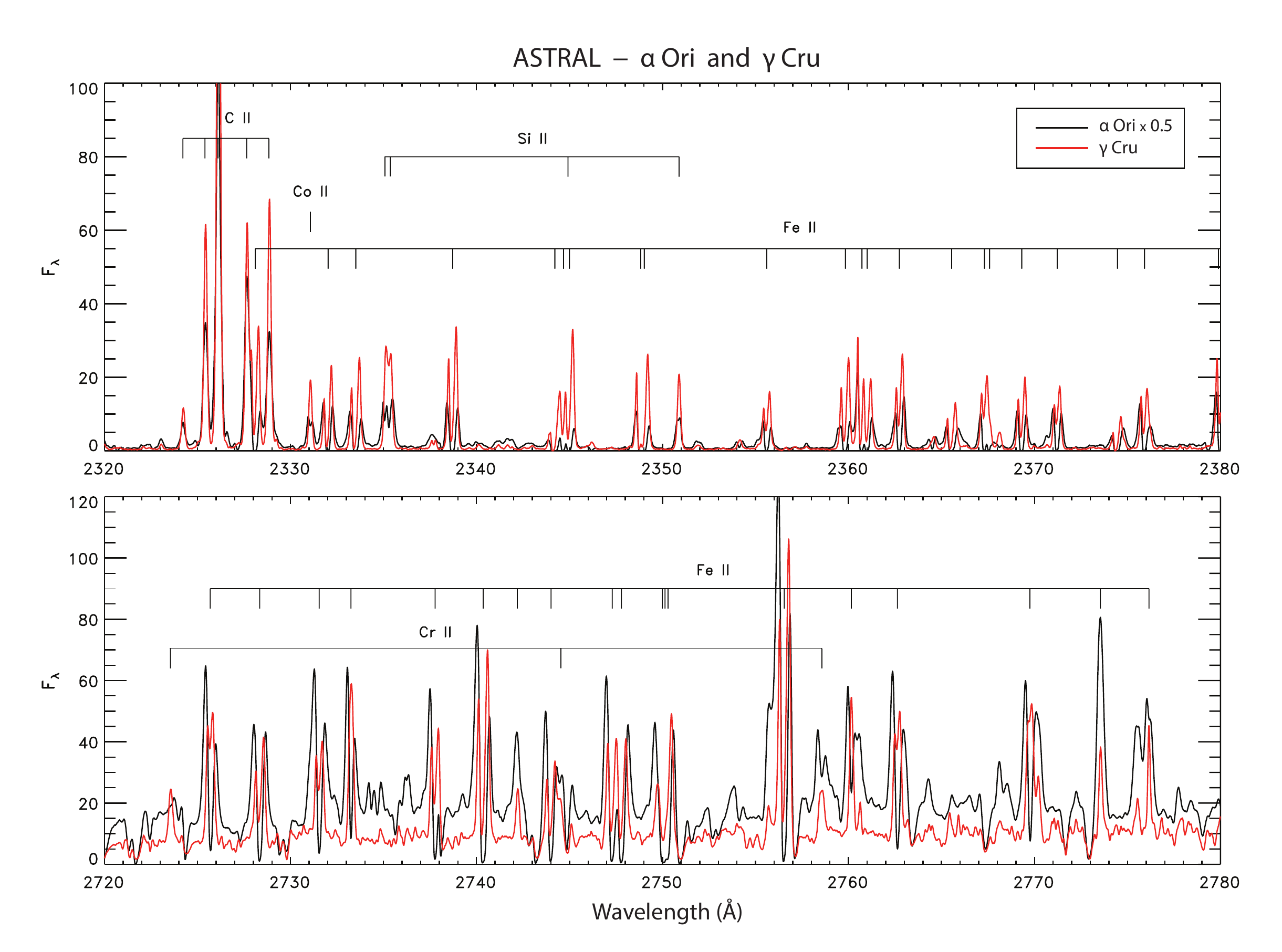}
\caption{The wealth of detail in these spectra is illustrated in these two zoomed-in panels, showing the 2320-2380  \AA\  and 2720-2780 \AA\ spectral regions. Selected emission lines are identified.  Flux is in units of 10$^{-13}$ ergs cm$^{-2}$ s$^{-1}$ \AA$^{-1}$.   \label{zoom}}
\end{center}
\end{figure*}

An overview of the spectra along with identifications of the major features are shown in Figure~\ref{zoom} and Appendix~\ref{plots}. Figure~\ref{zoom} zooms in to show an example of the details in the high resolution segments between 2320$-$2380 \AA\ and 2720$-$2780 \AA, while the Appendix shows summary plots of the entire combined spectra. The identifications of the features in the spectra are made using the Kurucz\footnote{http://kurucz.harvard.edu}  and NIST\footnote{http://www.nist.gov/pml/data/asd.cfm} line databases, in addition to  work on the \aboo\ UV Spectral Atlas \citep{Ayres86a, Ayres86b, Hinkle05a}, GHRS studies of \lvel\ \citep{Carpenter99a, Carpenter14a}, the \aori\ study \citep{Carpenter84a, Carpenter94c, Carpenter97a, Carpenter99a}, and the \gcru\ \citep{Carpenter88a, Carpenter95a} analysis. In addition to wavelength coincidence all expected lines are analyzed based on their intrinsic line strength, e.g., $gf-$value, and a determination if all the lines expected from a given energy level (lower if absorption, upper if emission) were observed in the covered wavelength range. For the emission lines characterized as fluorescent it is required in general that the exciting line and the excitation route is identified. Lines and excited levels above those populated in plasma with a temperature $\sim$20,000~K or lower were not considered unless a radiative excitation process was found to populate the level. In the following section  the major components observed in the spectra are discussed.

\begin{figure*}
\begin{center}
\includegraphics[width=\textwidth]{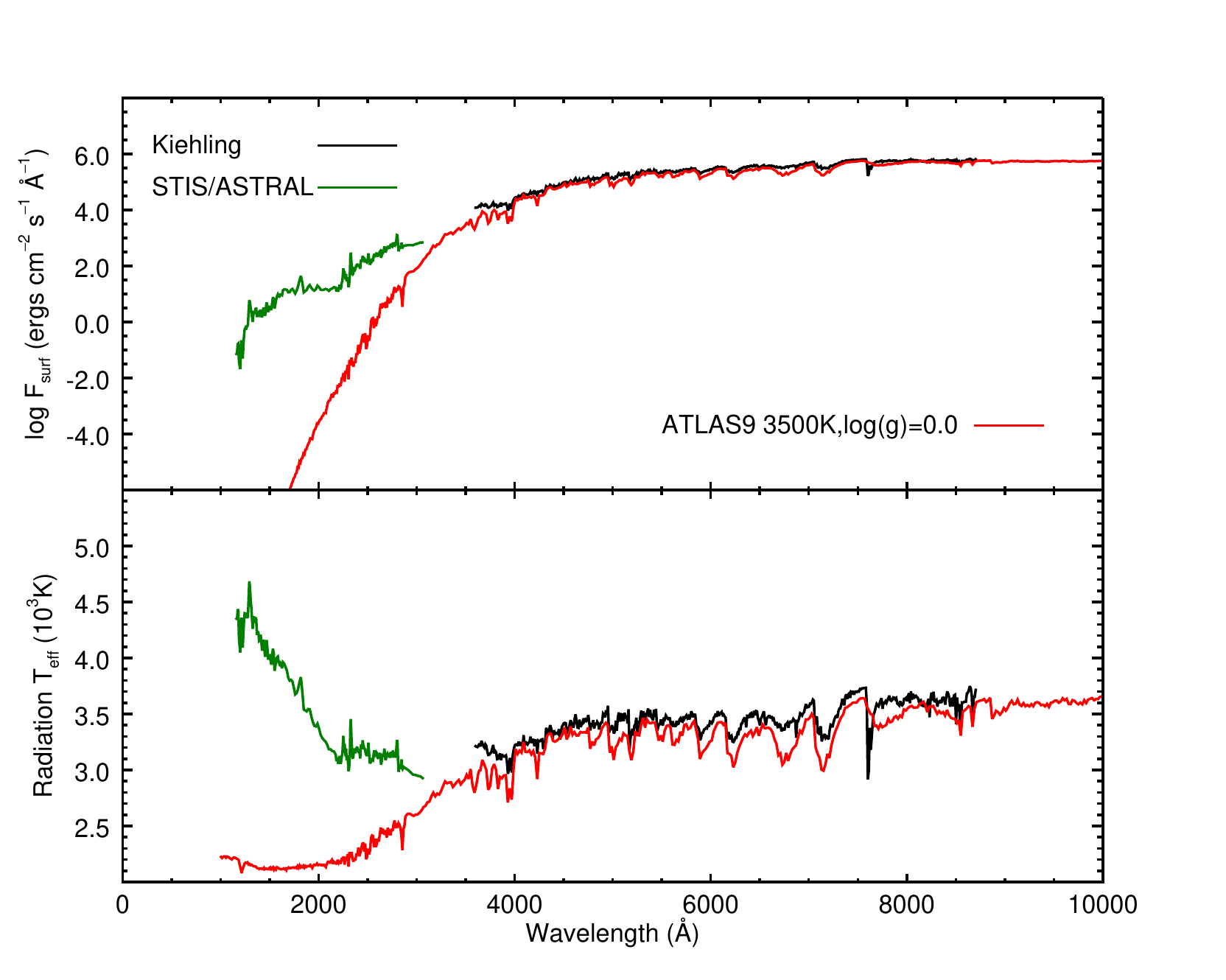}
\caption{Top: Comparison of the observed continuum surface flux distribution from \aori\ with a  modeled Kurucz LTE ATLAS9 synthetic continuum. Observed optical data (black) are from \citet{Kiehling87a}, the UV data are from HST/STIS (green), while the synthetic continua were computed with a standard Kurucz model (red). Orders of magnitude separate the observed UV from the photospheric models; clear evidence of continuum emission from the hotter chromospheric layers.  Bottom: The chromospheric temperature rise implied by the observed fluxes (black and green lines) is shown in this plot of the effective radiation temperature vs. wavelength.  The Kurucz model temperature vs. wavelength is also shown for comparison. \label{cont_aori} }
\end{center}
\end{figure*}

\begin{figure*}
\begin{center}
\includegraphics[width=\textwidth]{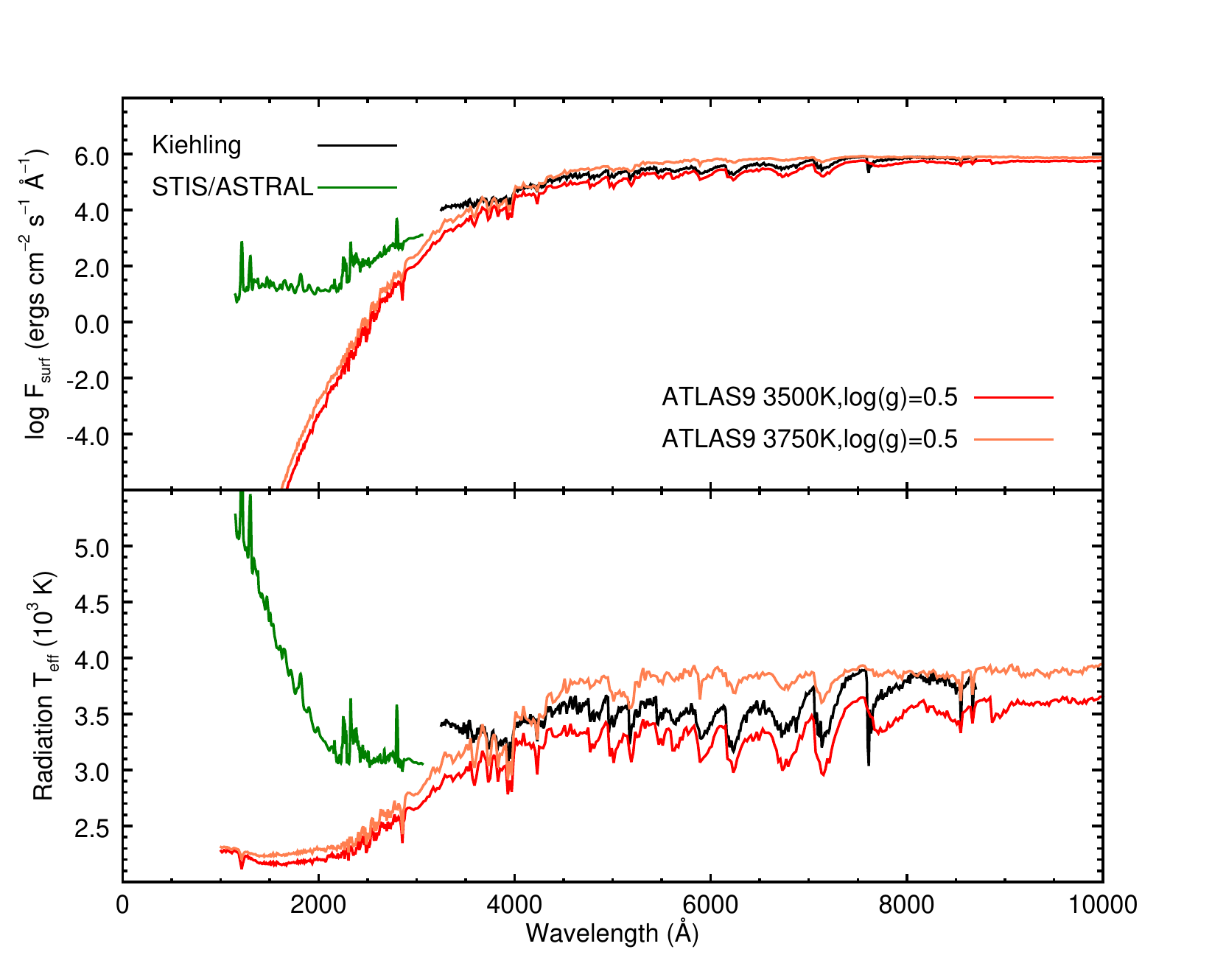}
\caption{Top: Continuum for \gcru\ shown as in the preceding figure for \aori. Modeled Kurucz LTE ATLAS9 flux (3500~K in red; 3750~K in orange ) compared to observed flux by \citet{Kiehling87a} in the optical and ASTRAL flux in the ultraviolet. Bottom: The chromospheric temperature rise implied by the FUV continuum in \gcru.  The Kurucz model temperature vs. wavelength is also shown for comparison. \label{cont_gcru}}
\end{center}
\end{figure*}

\subsection{The Chromospheric Continua }
These objects display a strong far-UV continuum whose flux is orders of magnitude brighter than expected from the photospheres of M giant or supergiant stars. The excess flux is thought to be formed in the chromosphere. Although it is visible in earlier \iue\ data, the chromospheric continuum emission was only understood to be present after GHRS observations of \aori\ were obtained and analyzed to reveal that the FUV spectrum was dominated by substantial circumstellar CO A-X band absorptions superposed over a chromospheric continuum, rather than just a collection of blended chromospheric emissions lines \citep{Carpenter94c}.  The FUV continuum has also been observed in the K5 supergiant \lvel\ \citep{Carpenter14a} but it is more uniform for \lvel\ due to the absence of the CO (A$-$X) absorption bands. The continua for \aori\ and \gcru\ are presented in Figures~\ref{cont_aori} and \ref{cont_gcru}, respectively, where the observed flux in the optical \citep{Kiehling87a} and ultraviolet (ASTRAL) wavelength regions are compared to a computed flux using an ATLAS9 model\footnote{http://www.oact.inaf.it/castelli/}. The calculations of the observed surface fluxes for \aori\ and \gcru\ are based on angular diameters of 43.8 mas \citep{Perrin04a} and  24.7  mas \citep{Glindemann01a} for \aori\ and \gcru, respectively. 

The observed and the modeled fluxes are in agreement for both objects down to $\sim$3500 \AA, but in the ultraviolet the fluxes diverge and differ by 5 to 6 orders of magnitude below 2000~\AA. \citet{Carpenter99a, Carpenter14a} compared the flux for \lvel\ with a  model where the temperature in the upper layers was not allowed to fall below a minimum temperature  of 3200~K ($T_{min}/T_{\it eff}$ = 0.8), that reproduced the flux level for \lvel\ down to 2000 \AA\ significantly better, but could not reproduce the flux at shorter wavelengths. The observed flux requires the presence of a temperature rise (as shown in Figures~\ref{cont_aori} and \ref{cont_gcru}) above the photospheric value and even above the 3200~K temperature minimum and thus that the origin of the flux is in the chromosphere. The \gcru\ and \aori\ continua show a similar behavior,  with an even larger flux discrepancy at ultraviolet wavelengths.  

\subsection{The Emission Spectrum}\label{emission}
The emission spectrum in \aori\ and \gcru\ is strongly influenced by the radiation of strong chromospheric features and the surface blackbody radiation. Many of the light elements with low ionization limits are ionized and show recombination spectra, while other  ions and molecules show strong signatures of selective excitation processes.  We provide details of our identifications and measurements of the emission lines in the ASTRAL spectra of \aori\ and \gcru\ in Appendix B, where Tables~\ref{tbl-7} through \ref{tbl-9} show, respectively, the measured properties of: 1) the fluorescent emission lines in \aori\ and \gcru, 2) the chromospheric and wind lines in \gcru, and 3) the chromospheric and wind lines in \aori. The following paragraphs briefly discuss the presence of some of the observed spectral features in the \aori\ and \gcru\ spectra.

\subsubsection{Atomic Emission}
\begin{figure*}[h]
\includegraphics[width=\textwidth]{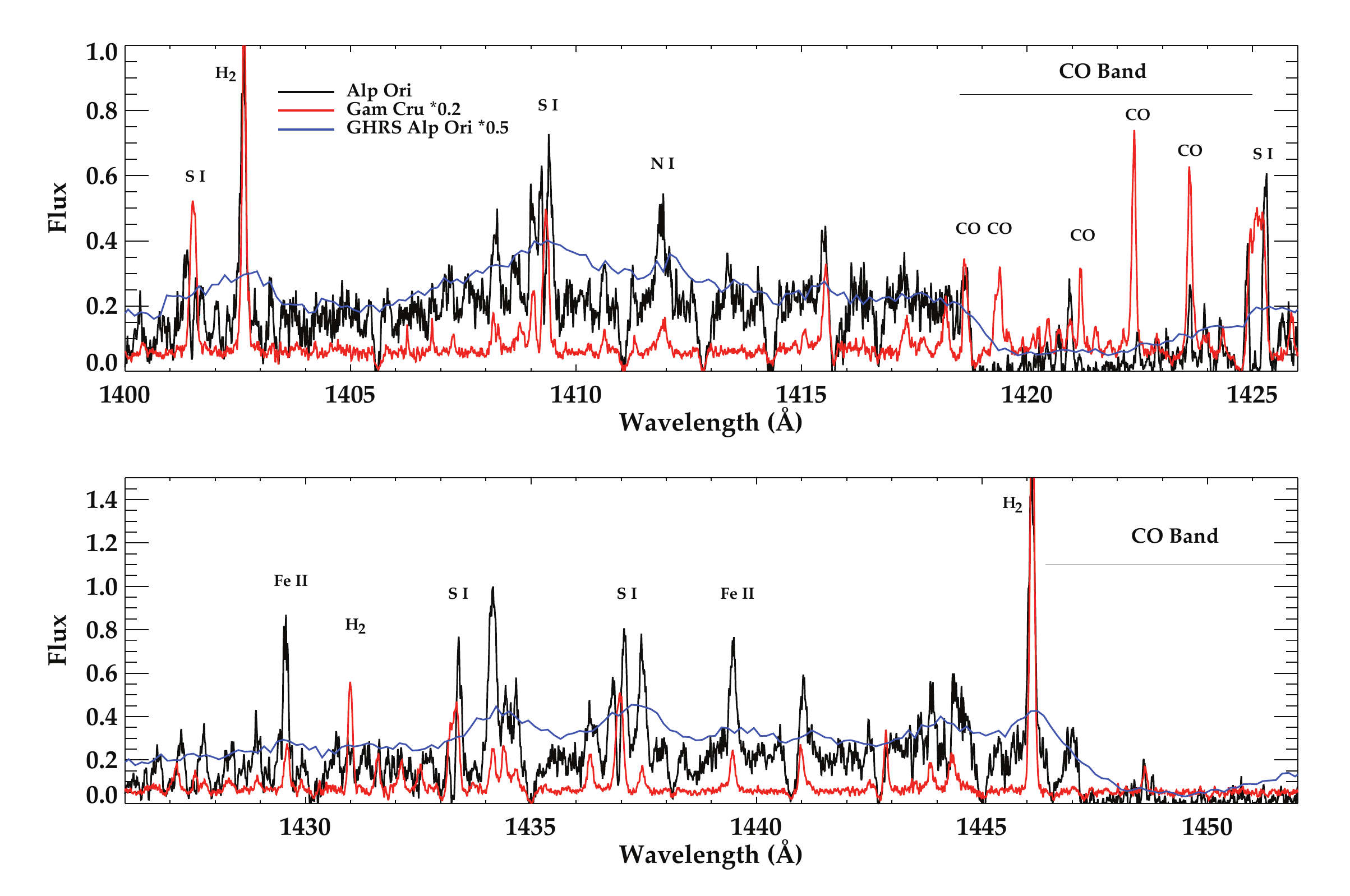}
\caption{The search for cool plasma in these two stars: the 1400$-$1452 \AA\ region of the ASTRAL \aori\ and \gcru\ spectra, with a lower resolution GHRS spectrum of \aori\ superposed to show  its circumstellar CO A$-$X absorption bands. Note the narrow fluorescent CO emission lines that are superposed over the much broader CO absorption bands. The CO absorption bands are signatures of cool material in the circumstellar shells of \aori, while the fluorescent CO emission in both stars is a signature of cool material embedded in their chromospheres. Fluxes are in units of 10$^{-13}$ ergs\,cm$^{-2}$\,s$^{-1}$\,\AA$^{-1}$. \label{Cool_Plasma}}
\end{figure*}

The ionization potential for neutral sulfur is close to the \lya\ transition energy, hence the hydrogen radiation populates numerous levels close to the S$^0$ ionization limit \citep{Tondello72a, Judge88a}. The cascades to the ground state produce a prominent \ion{S}{1} emission spectrum in both \aori\ and \gcru\ {(see Figure~\ref{Cool_Plasma}, Figure~\ref{Fig10} panel 3, and Figure~\ref{Fig11} panels 2, 3, and 5)} . Similarly, the \lya\ flux is  important for  C$^0$ and Si$^0$ that both can be ionized by the strong hydrogen flux, and is important for the formation of the \ion{C}{1}, \ion{C}{2}, \ion{Si}{1} and \ion{Si}{2} spectra. Neutral carbon can be ionized from the $^1$D (10,192 \cm) and $^1$S (21,648 \cm) terms and form {a prominent  \ion{C}{1} spectrum}, in particular, at wavelengths {below 2000~\AA\ (some examples can be seen in Figure~\ref{Fig10} panels 3 and 5 and in Figure~\ref{Fig11} panel 3)}. The neutral carbon spectrum is stronger in \gcru\ than \aori, indicating either a  weaker \lya\ or a lower carbon abundance in \aori,  but both spectra show a similar {disproportional strength to the} \ion{C}{1} \lm1657 and \ion{C}{1} \lm1993 {(see Figure~\ref{Fig10} panel 5 and Figure~\ref{Fig11} panel 3)}  that is caused by differences in the optical depth sensitivity \citep{Jordan67a}.  Neutral silicon shows a greater presence in \aori\ with stronger emission lines from the excited levels (E$_\mathrm{low}>$ 6000 \cm), but exclusively absorption or self-reversed line profiles for the resonance lines. 
  
Selective fluorescence processes  in cool giants and supergiants have previously been observed for the iron-group elements and  discussed by \citet{Carpenter94c,Carpenter95a}. These excitation channels  are enabled by strong chromospheric emission lines such as \lya\ and \lyb\ that produce observed fluorescence in, for example, \ion{O}{1}, \ion{Cr}{2}, \ion{Fe}{2}. The strong neutral oxygen spectrum observed  in both \aori\ and \gcru\ is  caused by the wavelength coincidence of  \ion{O}{1} \lm1026 and \lyb\ populating the 3d$^3$D state at 97,488 \cm. Cascades from this upper level produce the prominent \ion{O}{1} 1355 \AA\ (UV1), \ion{O}{1} triplet at 1300 \AA\ (UV2) and, \ion{O}{1} \lm1641 in both objects {(see Figure~\ref{Fig10} panels 2 and 5)}. 

The \lya\ driven \ion{Fe}{2} fluorescence is strong  in both \aori\  and \gcru. The rich \ion{Fe}{2} spectrum has multiple transitions close in wavelength ($<$3 \AA) to the strong \lya\ line, resulting in numerous fluorescent transitions from Fe$^+$ levels between 90,000$-$110,000 \cm\ (see Table~\ref{HI_FeII}). The strong \lya\ also redistributes the population in the Fe$^{+}$ ground configurations, enabling additional fluorescence processes  from  higher meta-stable levels (E$_\mathrm{low}>$ 20,000 \cm) that normally would not be populated in this environment. For example, \citet{Johansson84a} showed that the levels at 107,674 and 107,720 \cm\ can be excited by \lya.  Some decays from the pumped levels  fall outside our covered wavelengths but  produce observable secondary fluorescence noticeable in \gcru, for example \ion{Fe}{2} \lm\lm2521, 2527, and 2539.  

\begin{table}[h]
\caption{Fe$^+$ energy levels populated via  \lya\ \lm1215 wavelength coincidence \label{HI_FeII}} 
\begin{tabular}{ccccc} \hline
$E_\mathrm{u}$(cm$^{-1}$)  & $E_\mathrm{l}$(cm$^{-1}$) & \lm (\AA) & $\mid \delta \lambda \mid$ & log\,gf \\ \hline \hline
\phantom{a}90,067\tablenotemark{a}  &  7,955 &  1217.848 &  2.17 &  $-$1.49  \\  %(3P) 4p 4G_9/2
90,398   &    7,955 &  1212.966 &  2.71 &  $-$1.47 \\  %5p 4D_7/2
90,639   &    7,955 &  1215.852 &  0.18 &  $-$2.03\\   %5p 4D_5/2
90,899   &    8,680 &  1216.272 &  0.60 &  $-$2.35\\   %5p 4D_3/2
91,200   &    8,846 &  1214.285 &  1.39 &  $-$2.16\\   %5p 4D_1/2
90,301   &    7,955 &  1214.398 &  1.28 &  $-$2.23\\   %5p 6F_7/2
90,387   &    7,955 &  1213.132 &  2.54 &  $-$1.18 \\  %5p 6F_9/2
90,630   &    8,391 &  1215.983 &  0.31 &  $-$3.05\\   %4p 4S_3/2
90,781   &    7,955 &  1213.759 &  1.91 &  $-$1.17\\   %5p 4F_7/2
91,071   &    8,680 &  1213.738 &  1.94 &  $-$1.31\\   %5p 4F_5/2 
91,209   &    8,846 &  1214.150 &  1.52 &  $-$1.49\\   %5p 4F_3/2 
90,839   &    8,680 &  1217.152 &  1.48 &  $-$2.22\\   %4p 4P_1/2
90,901   &    8,680 &  1216.239 &  0.57 &  $-$2.32\\   %(5D) 5p 4P_5/2
91,048   &    8,846 &  1216.523 &  0.85 &  $-$2.30\\   % (3P) 5p 4P_3/2 
95,858   &  13,673 &  1216.769 &  1.10 &  $-$3.79\\
95,996   &  13,673 &  1214.735 &  0.94 &  $-$3.07\\
103,967 &  21,812 &  1217.205 &  1.53 &  $-$2.08\\
104,817 &  22,637 &  1216.847 &  1.17 &  $-$1.47\\
104,938 &  22,637 &  1215.058 &  0.62 &  $-$1.78\\
105,029 &  22,810 &  1216.275 &  0.60 &  $-$1.97\\
107,674 &  25,428 &  1215.873 &  0.20 &  $-$1.34\\
107,721 &  25,428 &  1215.183 &  0.49 &  $-$1.50\\ \hline 
\end{tabular}
\tablenotetext{a}{5p $^4$F$_{9/2}$ at 90,043 \cm\  is pumped  via energy level \\mixing with  5p $^4$G$_{9/2}$ 90,067 \cm.}
\end{table}

The Cr$^+$ lines reflect a  \lya\  fluorescence spectrum that is weaker than for \ion{Fe}{2}.  The majority of the strong pumping channels in Fe$^+$ originate from Fe$^+$ metastable states  7955 to 8846 \cm\ while the lowest energy levels in Cr$^+$ with \lya\ coincidence are slightly higher,  at $\sim$12,000 \cm. The levels at 94,749  and  94,656 \cm\ show a strong influence from \lya\ and to a lesser degree also the energy levels at  94,522, 94,365, 94,452, 94,363  and 102,725 \cm.   No secondary fluorescence, analogous to that observed from  Fe$^+$,  is observed from Cr$^+$ in the \gcru\ or \aori\ spectra.   

The existence and productivity of a fluorescent pumping mechanism is dependent not only on the wavelength coincidence but also on the strength and width of the exciting feature.  Other chromospheric features that are sufficiently strong and broad, and have important coincidences with strong lines of other elements, are C$^+$, O$^0$, Mg$^0$, Mg$^+$, Si$^+$  and, even Fe$^+$ - all of which provide weaker but similar excitation routes as \lya\ and \lyb. 

Some of the other iron-group ions show signs of fluorescence, including \ion{Fe}{1}, \ion{Mn}{1}, \ion{Ti}{2},  \ion{Ni}{2}, \ion{Co}{2}. Neutral iron displays strong wind absorption spectrum  from the ground configuration, especially in \aori. The strongest ground state transitions are also  present in the \gcru\ spectrum and Fe$^{0}$ seems to have a strong population even in the first excited configuration. Accidental wavelength coincidences between \ion{Fe}{1} transitions from the a$^5$F term and  the strong \ion{Mg}{2} \lm 2796.35, \ion{Fe}{2} \lm  2725.69 result in fluorescent transition  above 2600 \AA.  A single line as a result of \ion{Fe}{1} coincident with \ion{Mg}{1} \lm 2853 is present in the \aori\ spectrum at 2838 \AA.  

The \ion{O}{1} \lm\lm1304, 1305, 1306 (that are cascades from \lyb\ pumping) give, while weaker in \aori, a strong fluorescent CO spectrum (as discussed in Section~\ref{MOL}) in \gcru\ in addition to  \ion{P}{2} \lm1309.87. The latter is influenced  by the resonance wind lines in \aori.  The \ion{C}{2} \lm1335.73 produces the \ion{Cl}{1} \lm1351.6 line, a process originally reported by \citet{Shine83a}, and likely \ion{Fe}{2} \lm\lm2492,2303 in \gcru.  All the \ion{C}{2} pumped features are absent or very weak in the \aori\ spectrum which is consistent with the intrinsic strength of the singly ionized carbon spectrum. Magnesium and silicon transitions selectively excite a few levels for the iron-group elements.

Transitions from levels $<$90,000 \cm, which are below what can be excited by \lya, are observed in both \aori\ and \gcru\ in Ti$^+$, Cr$^+$, Fe$^+$, Ni$^+$ and Co$^+$. These levels are selectively populated by strong \ion{Fe}{2} transitions. Some of these processes involving strong \ion{Fe}{2} transitions are described by \citet{Carpenter88b} but the impact of the strong \ion{Fe}{2} chromospheric wind spectrum, and  the energy level density for the iron-group elements enable a large number of wavelength coincidences that produce numerous excitation channels. The levels sp x$^6$P (79,246, 79,285 and 79,331 \cm)  are examples of levels pumped by \ion{Fe}{2} UV9 between 1260 and 1275 \AA, to produce fluorescent lines at 1780 \AA, \ion{Fe}{2} UV 191. The \ion{Fe}{2} UV9 to UV191 photon conversion was first described by \citet{Hempe82a}. \citet{Johansson95a}  later discussed the presence of satellite lines close to the 1785, 1786 and 1788 lines, due to level mixing between w$^2$P and x$^6$P. No satellite lines are resolved  in the spectra of  \gcru\ or \aori. \cite{Hempe82a} report that almost every photon to the x$^6$P levels is redirected to UV 191 instead of being reemitted in UV9. 
 
  \begin{table}[h]
\caption{H$_2$ transitions with \lya\ \lm1215 wavelength coincidence and populated energy levels. \label{HI_H2}} 
\begin{tabular}{cccc} \hline
\lm (\AA) & ID & $E_\mathrm{l}$(cm$^{-1}$)  & $E_\mathrm{u}$ (cm$^{-1}$) \\ \hline \hline
 1214.465 &   H$_2$(3-1)R15 &        15,649 &        97,990  \\
 1214.781 &   H$_2$(4-3)P5\phantom{5}   &        13,265 &        95,584  \\
 1215.727 &   H$_2$(1-2)R6\phantom{5}   &        10,261 &        92,516  \\
 1216.070 &   H$_2$(1-2)P5\phantom{5}   &       \phantom{1}9,654 &        91,886  \\
 1217.205 &   H$_2$(0-2)R0\phantom{5}   &       \phantom{1}8,086 &        90,242  \\
 1217.643 &   H$_2$(0-2)R1\phantom{5}   &       \phantom{1}8,193 &        90,319  \\
 1217.904 &   H$_2$(2-1)P13 &        13,191 &        95,299  \\ \hline
\end{tabular}
\end{table}

\subsubsection{Molecular Emission\label{MOL}}
Molecular emission is observed  in the form of fluoresced CO and  \h2\ emission, as seen in the K-giant stars \aboo\ (K1.5~III) and $\alpha$~Tau (K5~III).  Molecular hydrogen is the most abundant species in the interstellar medium and is fundamental for the formation of complex molecules and dust.   In most systems the temperature limits the population to the lowest  $J$-states in the ground vibrational level.  This results in observable transitions only in the far-UV in addition to rotational lines in the infrared. The \h2\ energy levels populated by coincidence with \lya\ are shown in Table \ref{HI_H2}.  Carbon monoxide is like \h2\ an abundant interstellar molecule with relatively simple spectroscopic properties.  The energy state population in carbon monoxide is restricted to its lowest vibrational states in cool temperature environments and the chromospheric conditions enable population of higher states and hence pumped transitions in our observed wavelength interval.  The most prominent spectrum in the UV is the fourth positive system, A $^1\Pi$$-$X $^1\Sigma^+$, with transitions between 1280 and 2800 \AA, however, our identifications are exclusively between 1330 and 1554 \AA. Fluorescent carbon monoxide has previously been observed in \aboo\ \citep{Ayres81a, Ayres99a} and $\alpha$ Tau \citep{McMurry00a}.

The molecular fluorescence is stronger in \gcru\ than for \aori. The \h2\ and CO spectra are powered by \lya\ and \ion{O}{1} \lm\lm1302, 1304, 1306, respectively. The large number of energy levels and their closeness in energy result in several lines in both \h2\ and CO with wavelengths close to the pumping transitions. Other pumped channels for CO  may include \lya\ and \ion{C}{1} $\lambda$1656$-$1658, but fluorescence fed by these transitions are not identified in either \gcru\ or \aori.  These CO and \h2\ fluorescence spectra are believed to be formed in the objects' outer atmosphere and by studying them, information regarding the physical structure of the emitting gas and the radiative source from the inner parts are gained.  Alpha Boo and $\alpha$~Tau do not show the strong far-ultraviolet continuum as observed for  \aori\ and  \gcru, which is consistent with the greater column densities of {the cooler and more massive M-star chromospheres,} where $\tau$=1 surfaces are shifted to larger radii as the chromospheric continuum is formed further out at a greater temperature and producing a larger surface flux.

The CO  fluorescence in \aori\ is more difficult to analyze, due to the circumstellar shell engulfing the object. The circumstellar absorption is strong and makes the identification of some of the fluorescent CO lines uncertain. 
 
We have measured the fluorescent atomic and molecular emission line features  and present the line identifications, wavelengths, integrated and surface fluxes, and apparent velocity relative to the stellar photospheric radial velocity in Table \ref{tbl-7} in the Appendix. 

\subsection{Diagnostics of Cool Plasma}
The spectra of both stars contain numerous diagnostics of cool plasma, including the atomic emission as showed in Figure~\ref{Cool_Plasma} from chromospheric \ion{S}{1} and \ion{N}{1} as well as fluorescent molecular \h2\ and CO emission from low temperature pockets of material immersed in the chromosphere. In addition, the spectrum of \aori\ contains broad CO A$-$X bands produced in its circumstellar shells \citep{Bernat78a, Carpenter94c}. 

Strong absorption bands of the CO (A$-$X) system are observed in the \aori\ far-ultraviolet spectrum, but there is no evidence of their presence in the \lvel\ or \gcru\ spectra.  However, there are clear bands of the OH molecule in both of the M stars, which show a major band-head of OH at 2811 \AA.  

\subsection{Diagnostics of Hot Plasma}
\begin{figure}[h]
\includegraphics[width=\columnwidth]{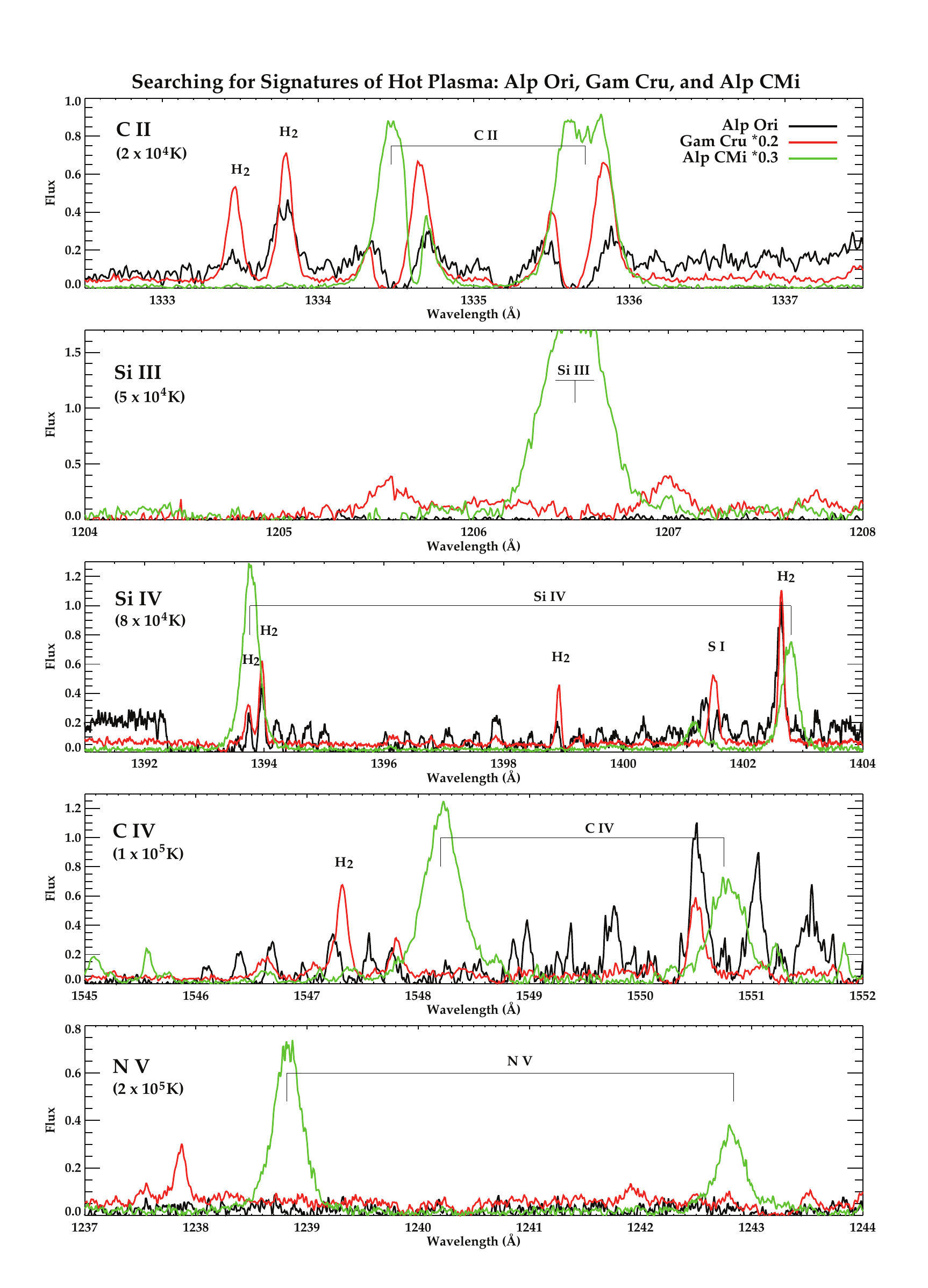}
\caption{A search for evidence of hot plasma in \aori\  and \gcru\  for the regions around five lines indicative of temperatures ranging from 2$\times$10$^{4}$~K to 2$\times$10$^{5}$~K.  The approximate formation temperature of each line is given in parenthesis for the section of spectral diagnostics.  Flux is in units of 10$^{-13}$ ergs cm$^{-2}$ s$^{-1}$ \AA$^{-1}$ \label {HotPlasma} }
\end{figure}

A search was conducted  for  spectral features from \ion{C}{2}, \ion{C}{4}, \ion{Si}{3},  \ion{Si}{4}  and  \ion{N}{5}  as indicators of hot plasma in the envelopes of   \gcru\ and \aori\ as showed in Figure~\ref{HotPlasma}. The \aori\ and \gcru\ spectra are compared to a reference spectrum of coronal-type  F5~IV star $\alpha$~CMi. All the hot plasma diagnostics are present in the $\alpha$~CMi spectrum. The \ion{C}{2} \lm 1335 is the hottest diagnostic ($\sim$2$\times$10$^{4}$~K) seen in these two M-stars, \gcru\ and \aori.

\section{Comparison with spectrum of $\alpha$ Boo}
\begin{figure}[h]
\includegraphics[width=\columnwidth]{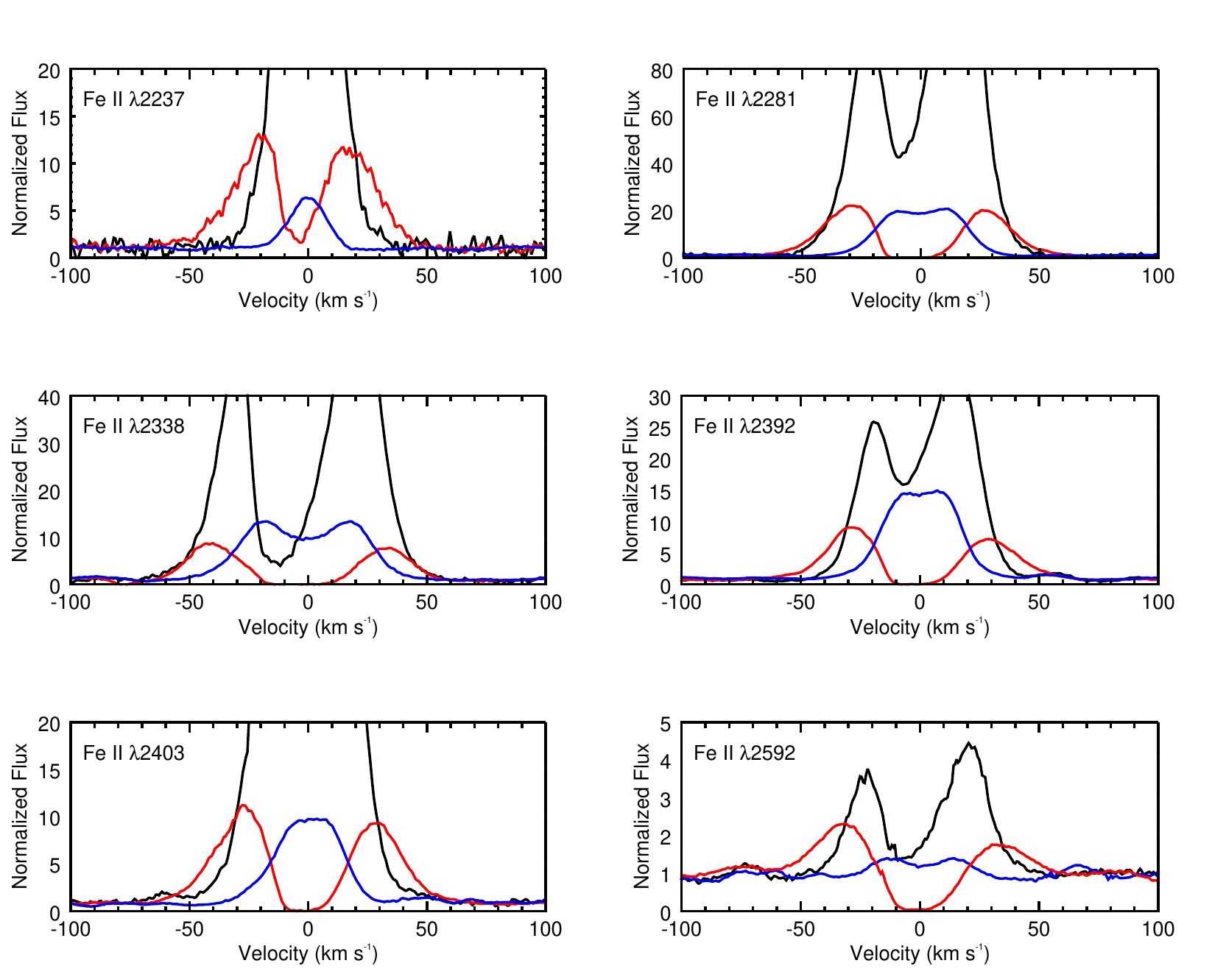}
\caption{Spectral comparison of the \ion{Fe}{2} line profiles observed in \gcru\ (black), \aori\ (red) and \aboo\ (blue). Plots are normalized to continuum 50-100 \kms\ from line center. \label {FeComp} }
%\end{figure}

%\begin{figure}[h]
\includegraphics[width=\columnwidth]{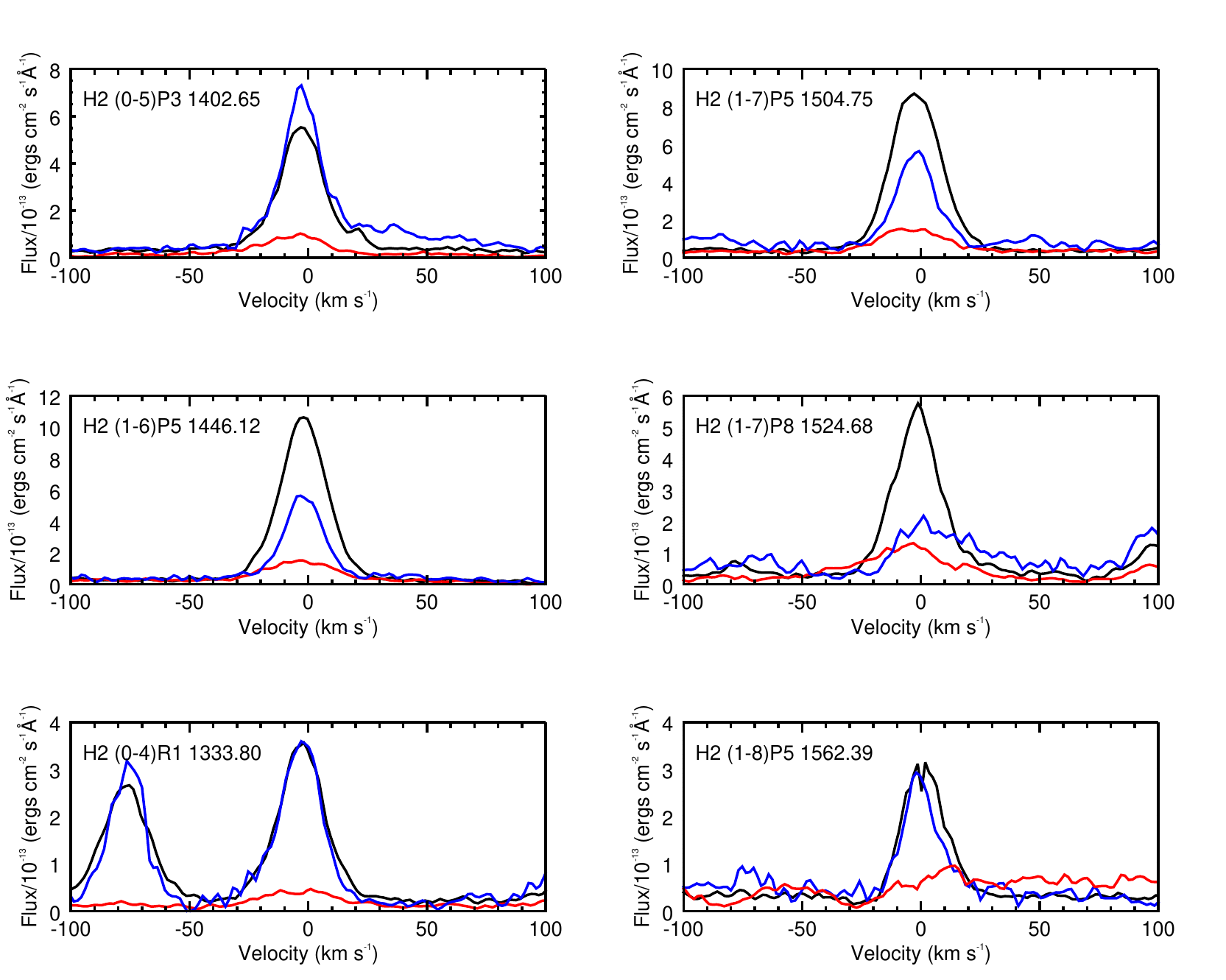}
\caption{Spectral comparison of the \h2\ line profiles observed in \gcru\ (black), \aori\ (red) and \aboo\ (blue) .\label {h2Comp} }
\end{figure}

\begin{figure}[h]
\includegraphics[width=\columnwidth]{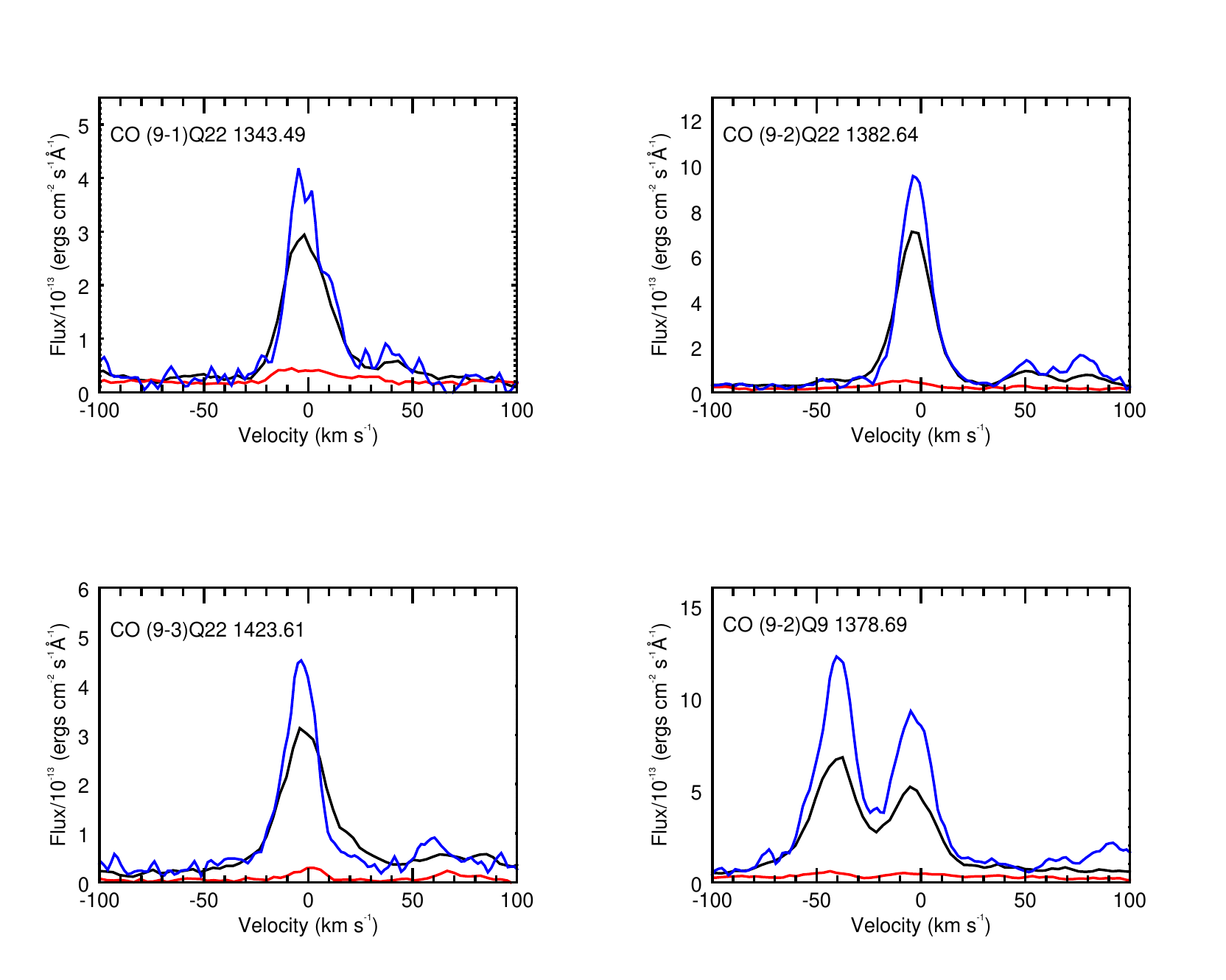}
\caption{Spectral comparison of the CO line profiles observed in \gcru\ (black), \aori\ (red) and \aboo\ (blue).  \label {COComp} }
\end{figure}

In Figures \ref{FeComp} through \ref{COComp}, we compare profiles for a set of atomic (\ion{Fe}{2}) emission lines and two sets of molecular,  \h2\ and CO, emission lines in the two M-stars \aori\ and \gcru\ to those in the prototypical K1.5 giant \aboo.  In Table~\ref{cf_surf_fluxes} we show the surface fluxes for the chromospheric emission components of these lines as a proxy of the magnetic activity levels (Fe~II) and the efficiency of the fluorescence mechanisms (\h2\ and CO) in the outer atmosphere of each star.

\begin{deluxetable*}{lcccccc}
\tablecolumns{7}
\tabletypesize{\scriptsize}
%\rotate
\tablecaption{Comparison of Selected Surface Fluxes from \aori,  \gcru, and \aboo\  Integrated Line Surface Fluxes in units of ergs cm$^{-2}$ sec$^{-1}$  \label{cf_surf_fluxes}}
\tablewidth{0pt}
\tablehead{\colhead{Lines} & \colhead{Wavelength} & \colhead{\aori} &\colhead{\gcru} &\colhead{\aboo}\  \\
& (\AA, vac.) & F$_\mathrm{surf}$ & F$_\mathrm{surf}$  & F$_\mathrm{surf}$ }
\startdata
Fe II &  2237.4  & 83.7 &  98.8 & 105.7 \\
& 2280.6  & 800.4 &  2566.8 & 2362.2 \\
& 2338.7  & 927.7 &  3160.6 & 3306.0 \\
& 2392.2  & 1305.0 &  3043.1 & 1955.6 \\
& 2403.3  & 798.4 &  1235.9 & 2686.3 \\
& 2592.3  & 391.3 &  704.2 & 356.0 \\ [2mm]
\h2\ &  1402.65  & 1.7  & 22.4   &  20.2  \\ 
&  1446.12  & 3.3  & 50.6  & 21.0  \\ 
&  1333.80  & 0.9  & 15.3   & 13.6  \\ 
&  1504.76  & 3.1  & 45.9  & 19.5  \\ 
&  1524.66  &  2.8 & 25.4  &  7.6 \\ 
&  1562.39  &  0.3 &  12.8  &  8.6 \\ [2mm]
CO &  1343.49  & 0.5 & 13.4 & 14.9 \\
&  1423.61  & 0.3  & 14.3 &  14.6 \\
&  1382.64  &  0.7 & 27.5 &  30.7 \\
&  1378.69  & \nodata & 24.5 & 37.1  \\
\enddata
%\tablecomments{any needed notes can be put here. }
\end{deluxetable*}

The lines of \ion{Fe}{2} in Figure \ref{FeComp} are chosen to represent a wide range in profiles, but it is clear that the strength of the overlying wind absorptions are always much stronger in a given line in \aori\ than in the other two stars, while the lines in \aboo\ are both narrower and weaker in their chromospheric emission and have very mild, if any, wind absorptions superposed. If we examine the \ion{Fe}{2} surface fluxes in Table~\ref{cf_surf_fluxes} of these lines in the three stars we see that generally the fluxes are substantially larger in the two giant stars \gcru\ and \aboo\ vs. the uniformly weaker lines in the supergiant \aori.

The lines of \h2\ in Figure \ref{h2Comp} are generally similar in the two giant stars \aboo\  (K1.5) and \gcru\ (M3.4),  though the latter has systematically stronger surface fluxes.  The lines in the M-supergiant \aori\ (M2 Iab) in contrast are considerably weaker in both observed flux (at earth) and surface flux than the lines in either of the two giants.  

The CO lines in Figure \ref{COComp} are also uniformly much stronger in the two giant-stars than in the M-supergiant, in which the CO emission is very weak, which may be due in part to substantial CO absorption in the circumstellar shell around \aori.  The surface fluxes in the two giants are very similar, while in \aori\ the fluxes are very small in three of the lines and unmeasurable in the fourth line. 

\section{Summary}

We have presented a detailed overview of the definitive STIS UV high-resolution spectra of two evolved M-stars, M2Iab supergiant \aori\ and the M3.4 giant \gcru, obtained as part of the \hst\ Treasury Program ``Advanced Spectral Library (ASTRAL) Project: Cool Stars"  to facilitate the use of this important archival dataset by future investigators.  This study provides identifications of the significant atomic and molecular emission and absorption features and discusses the characteristics of the line spectra.  The strong chromospheric continua seen in the far- and mid-UV are quantified and the merger of these continua with the photospheric continua seen at longer wavelengths is discussed.  The fluorescent processes responsible for a large portion of the emission line spectrum, the characteristics of the stellar winds, and the available diagnostics for hot and cool plasmas are summarized.  Finally, we have discussed differences between these spectra and existing UV spectra of the prototypical K1.5~III giant \aboo\ to illustrate the changing nature of the spectra as one moves from the K-giants into the cooler M-stars. 

Other recent work on \aori\ may be useful in understanding and putting these UV observations in context.  Perhaps most significant is {the} \citet{Petit13} report of a detection of a surface magnetic field in \aori\ using a time-series of six circularly polarized spectra obtained using the NARVAL spectropolarimeter at Telescope Bernard Lyot (Pic du Midi Observatory (F)), between March and April 2010. Zeeman signatures were repeatedly detected in cross-correlation profiles, corresponding to a longitudinal component of about 1 G, and showing a smooth increase of the longitudinal field from 0.5 to 1.5 G that correlated with radial velocity fluctuations. The correlation of Stokes V with radial velocity fluctuations, and their red-shift of about 9~\kms\ with respect to the Stokes I profiles, suggest that the observed magnetic elements may be concentrated in the sinking components of the convective flows. \cite{Tessore17} followed this up by investigating whether the magnetic field found on \aori\ was typical of red-supergiants or not and found that the two red supergiant stars in their sample, CE Tau and $\mu$ Cep, display magnetic fields very similar to that of \aori. They further reported that their non-detection of a magnetic field on the post-RSG star $\rho$~Cas suggests that the magnetic field disappears, or at least becomes undetectable with present methods, at later evolutionary stages and that their  analysis of $\alpha$$^1$ Her supports the proposed reclassification of the star as an M-type asymptotic giant branch star. \cite{Wood16} discuss constraints on the winds and astrospheres of Red Giant Stars provided by {\it{HST}} observations, focussing on spectra of the Mg II h and k lines near 2800 \AA. They studied stellar chromospheric emission, winds, and astrospheric absorption, concentrating on spectral types between K1.5 III and M5 III (noncoronal stars that possess strong, chromospheric winds). They found a very tight relation between \ion{Mg}{2} surface flux and photospheric temperature, supporting the notion that all K2-M5 III stars are emitting at a basal flux level. Wind velocities are generally found to decrease with spectral type, from $\sim$40~\kms\ at K1.5 III to $\sim$20~\kms\ at M5 III. {The power law index in the \cite{Wood16} empirical relation F$_{Mg II}$ $\sim$ T$_{eff}^{9.1}$ is remarkably close to the index of a theoretical relation derived by \cite{Bohn84} linking the flux of acoustic power F$_{acoustic}$ and the effective temperature: F$_{acoustic}$ $\sim$ T$_{eff}^{9.75}$. This makes it highly likely that the basal flux is indeed related to acoustic power, rather than magnetic effects (we thank the anonymous referee for pointing this out to us).}

\acknowledgments

Support for Program number HST-GO-12278.05-A was provided by NASA through a grant from the Space Telescope Science Institute, which is operated by the Association of Universities for Research in Astronomy, Incorporated, under NASA contract NAS5-26555.

%{\it Facilities:} \facility{\hst\ (STIS)}, \facility{\hst\ (GHRS)}.

\bibliographystyle{apj}
\bibliography{master_100418}

\clearpage

\appendix	
\section{Full-Spectrum Plots \label{plots}}

\begin{figure}[b]
\begin{center}
\includegraphics[width=13cm]{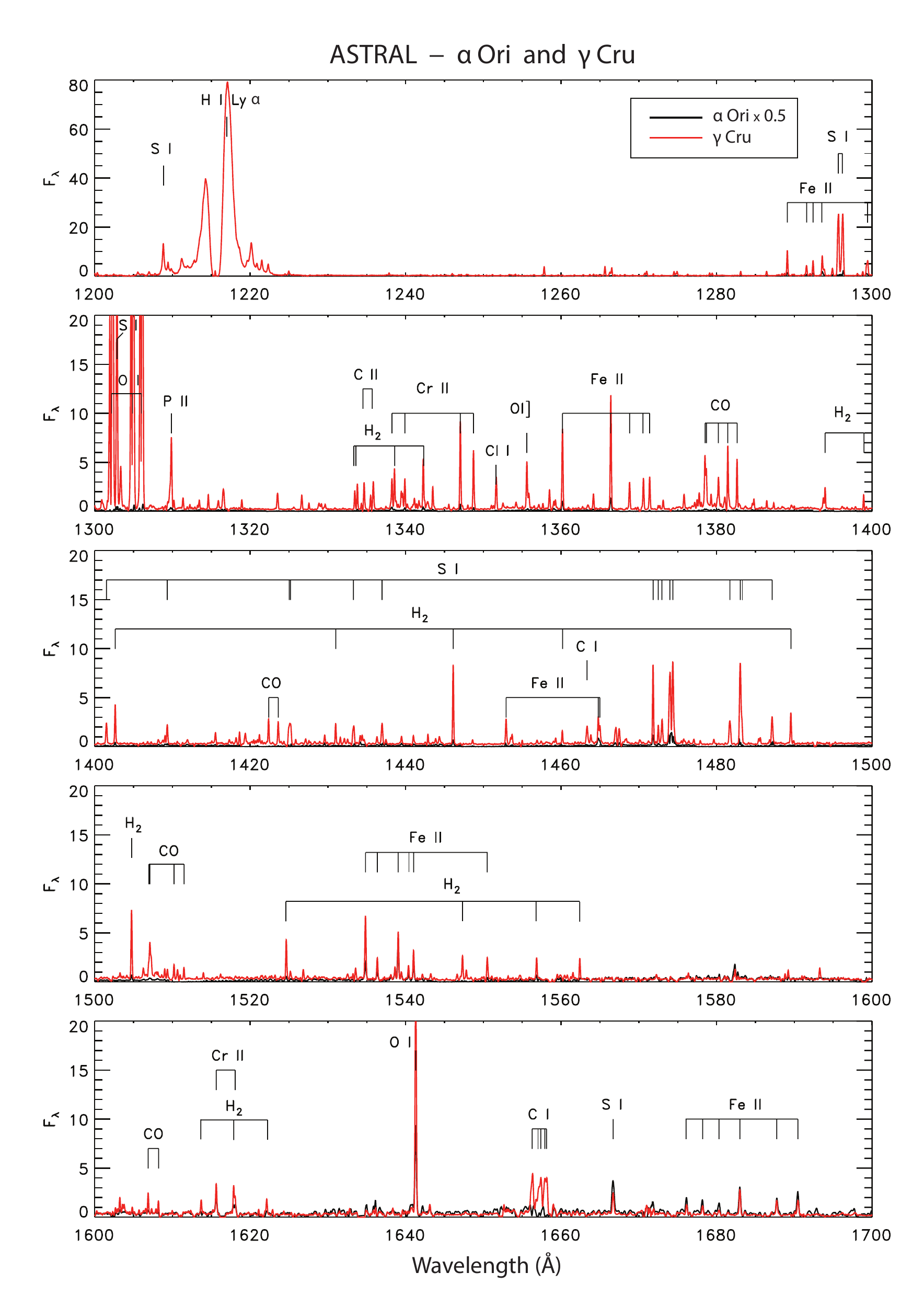}
\caption{The 1200$-$1700 \AA\ spectral region. Selected emission lines are identified.  Flux is in units of 10$^{-13}$ ergs cm$^{-2}$ sec$^{-1}$ \AA$^{-1}$.  \label{Fig10} }
\end{center}
\end{figure}

\begin{figure*}
\begin{center}
\includegraphics[width=13cm]{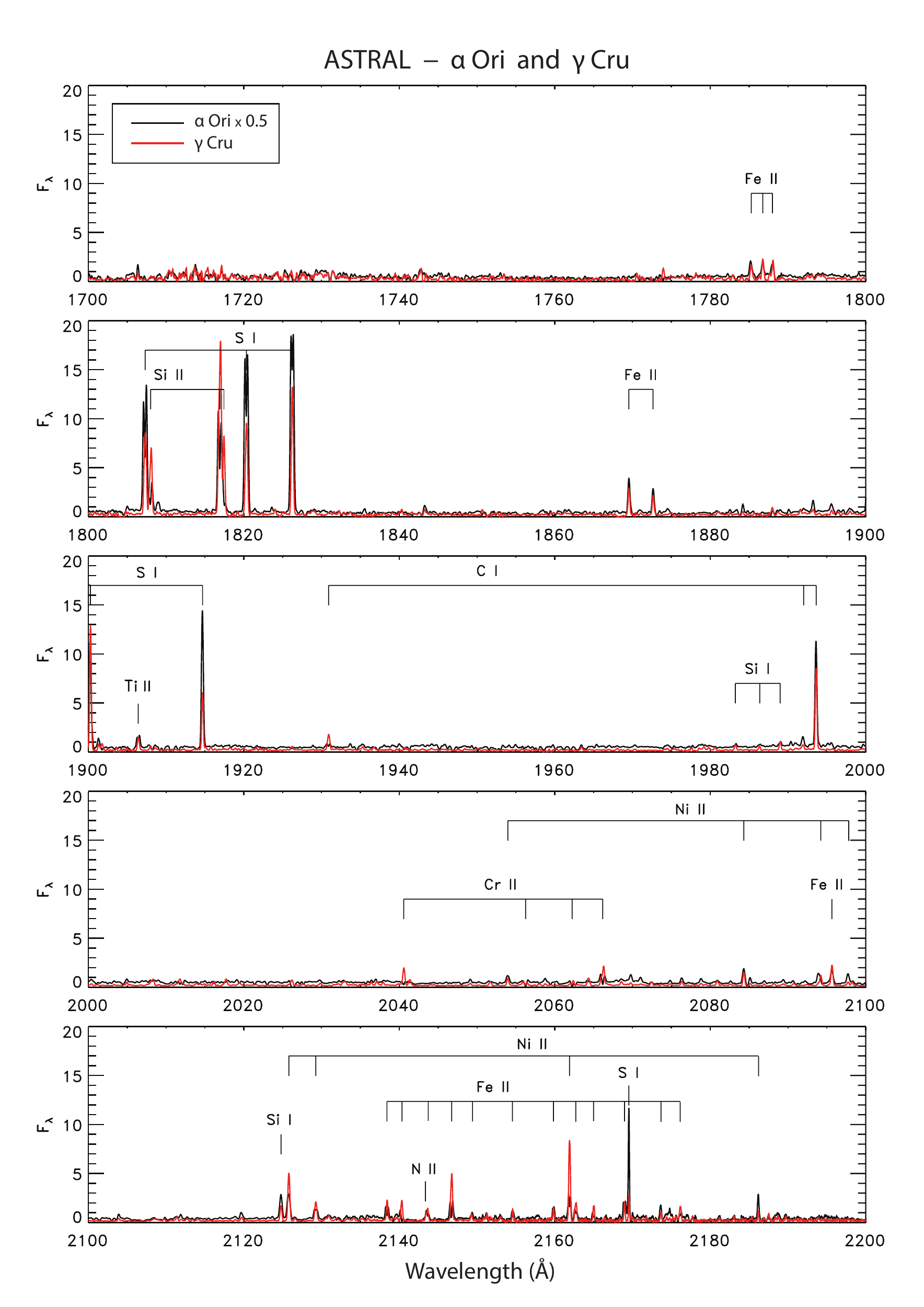}
\caption{The 1700$-$2200 \AA\ spectral  region is shown here. Selected emission lines are identified.  Flux is in units of 10$^{-13}$ ergs cm$^{-2}$ sec$^{-1}$ \AA$^{-1}$.\label{Fig11}}
\end{center}
\end{figure*}

\begin{figure*}
\begin{center}
\includegraphics[width=13cm]{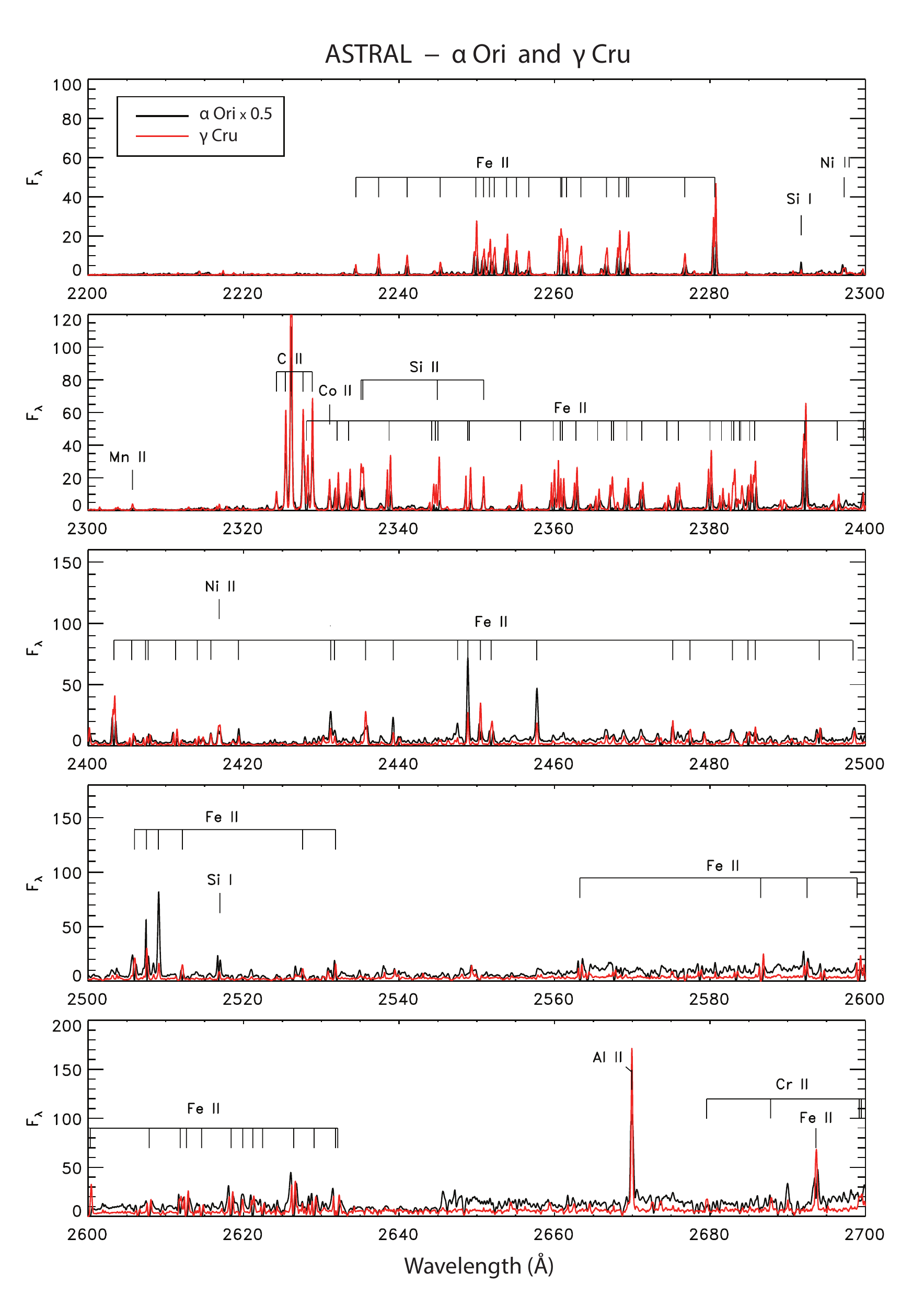}
\caption{The 2200$-$2700 \AA\ spectral region is shown here. Selected emission lines are identified.  Flux is in units of 10$^{-13}$ ergs cm$^{-2}$ sec$^{-1}$ \AA$^{-1}$.   \label{Fig12}}
\end{center}
\end{figure*}

\begin{figure*}
\begin{center}
\includegraphics[width=13cm]{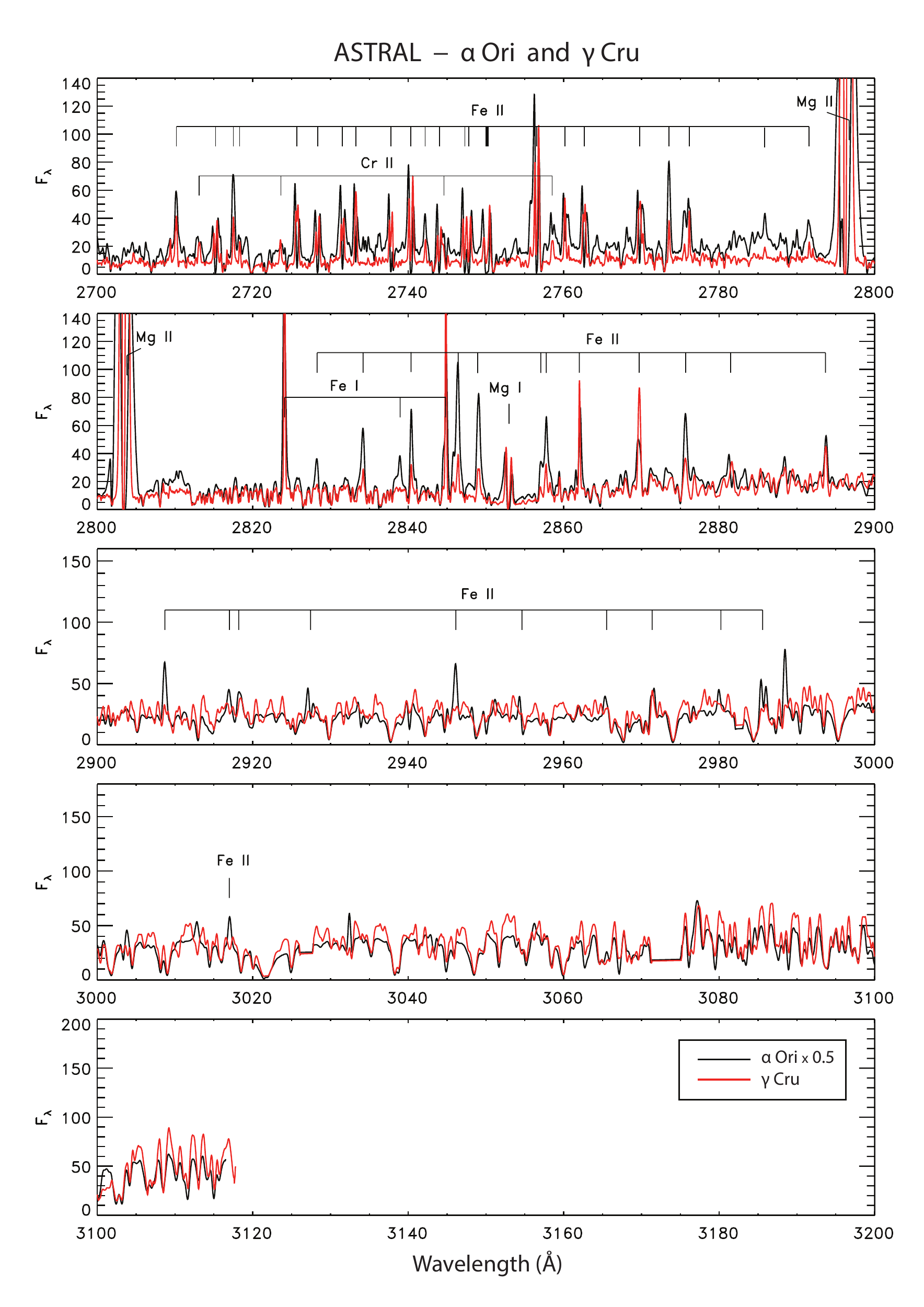}
\caption{The 2700$-$3120 \AA\ spectral region is shown here. Selected emission lines are identified.  Line Fluxes are in units of  10$^{-13}$ ergs cm$^{-2}$ sec$^{-1}$ \AA$^{-1}$.   \label{Fig13}}
 
\end{center}
\end{figure*}

Figures~\ref{Fig10} to \ref{Fig13} show the full ASTRAL spectra of the M stars \gcru\ and \aori.

\clearpage

\clearpage

\clearpage

\clearpage

\section{Detailed Tables: Emission Line Measurements \label{LongTables}}

Tables~\ref{tbl-7} through \ref{tbl-9} show, respectively, the measured properties of: 1) the fluorescent emission lines in \aori\ and \gcru, 2) the chromospheric and wind lines in \gcru, and 3) the chromospheric and wind lines in \aori.  The radial velocities (RV's) shown in these Tables are relative to the photosphere (stellar radial velocity of 20.6 km s$^{-1}$ for \gcru\  and 21.9 for \aori\ have been subtracted from the measured values). We show in the tables the most up-to-date laboratory atomic and molecular line data in the literature, taken primarily from the Kurucz\footnote{http://kurucz.harvard.edu}  and NIST\footnote{http://www.nist.gov/pml/data/asd.cfm} line databases.  \\
%
%LaTex environment for Table 7, version 6/26/18
%************************************
%\startlongtable
\begin{deluxetable*}{lcrr|rcrr|rcrr}[b]
\tablecolumns{12}
\footnotesize
\tablecaption{Properties of Fluorescent Emission Lines in STIS Spectra. {F$_\mathrm{int}$ is the integrated line flux observed at Earth, while F$_\mathrm{surf}$ is the integrated line flux at the surface of the star.} F$_\mathrm{int}$ is in units of 10$^{-15}$ ergs cm$^{-2}$ sec$^{-1}$; F$_\mathrm{surf}$ is in units of ergs cm$^{-2}$ sec$^{-1}$   \label{tbl-7}}
\tablehead{
\colhead{} & \colhead{} & \colhead{} & \colhead{} & \multicolumn{4}{c}{\gcru} &  \multicolumn{4}{c}{\aori} \\
\colhead{$\lambda_{lab}$ } & \colhead{Spectrum}  & \colhead{$\log gf$ } & \colhead{E$_{up}$} &  
\colhead{RV}  &\colhead{FWHM} & \colhead{F$_\mathrm{int}$}  & \colhead{F$_\mathrm{surf}$}  & 
\colhead{RV}  &\colhead{FWHM} & \colhead{F$_\mathrm{int}$}  & \colhead{F$_\mathrm{surf}$} \\
\colhead{(\AA, vac.) } & \colhead{} & \colhead{ }  & \colhead{(\cm)} &  
\colhead{(\kms)}  &\colhead{(\kms)} & \colhead{ }  & \colhead{ }  & 
\colhead{(\kms)}  &\colhead{(\kms)} & \colhead{ }  & \colhead{ }
}
\startdata 
1196.670 & Fe II & $-$1.09 & 104,817 & $-$0.4 & 38 & 10.6 & 5.0 & \nodata & \nodata & \nodata & \nodata \\
1199.236 & Fe II & \phantom{$-$}0.02 & 104,817 & $-$2.5 & 37 & 59.9 & 28.4 & \nodata & \nodata & \nodata & \nodata \\
1199.671 & Fe II & $-$0.11 & 104,938 & 8.5 & 35 & 37.9 & 18.0 & \nodata & \nodata & \nodata & \nodata \\
1202.454 & H2 1-2R3 & $-$0.11 & 91,886 & $-$1.7 & 39 & 10.8 & 5.1 & \nodata & \nodata & \nodata & \nodata \\
1205.606 & Fe II & $-$2.67 & 90,901 & $-$10.0 & 60 & 32.0 & 15.1 & \nodata & \nodata & \nodata & \nodata \\
1207.015 & S I & \nodata & 82,849 & $-$5.0 & 42 & 23.9 & 11.3 & \nodata & \nodata & \nodata & \nodata \\
1207.760 & S I & \nodata & 82,849 & $-$1.0 & 26 & 6.9 & 3.3 & \nodata & \nodata & \nodata & \nodata \\
1208.860 & S I & \nodata & 82,723 & $-$8.6 & 69 & 344.4 & 162.2 & \nodata & \nodata & \nodata & \nodata \\
1209.431 & Fe II & $-$1.79 & 90,639 & $-$3.1 & 88 & 139.7 & 65.8 & \nodata & \nodata & \nodata & \nodata \\
1209.829 & Fe II & $-$4.37 & 91,048 & 10.3 & 61 & 46.8 & 22.0 & \nodata & \nodata & \nodata & \nodata \\
1211.220 & S I & \nodata & 82,562 & $-$3.1 & 97 & 216.6 & 101.8 & \nodata & \nodata & \nodata & \nodata \\
1212.820 & S I & \nodata & 82,850 & $-$10.3 & 64 & 37.8 & 17.8 & \nodata & \nodata & \nodata & \nodata \\
1220.162 & S I & \nodata & 82,352 & $-$3.2 & 99 & 421.2 & 196.7 & \nodata & \nodata & \nodata & \nodata \\
1220.872 & Fe II & $-$2.01 & 90,301 & $-$3.2 & 54 & 54.7 & 25.5 & \nodata & \nodata & \nodata & \nodata \\
1221.492 & Cr II & $-$0.60 & 94,363 & 1.7 & 50 & 92.1 & 43.0 & \nodata & \nodata & \nodata & \nodata \\
1222.330 & S I & \nodata & 82,208 & $-$2.3 & 47 & 67.2 & 31.3 & \nodata & \nodata & \nodata & \nodata \\
1224.990 & S I & \nodata & 82,208 & $-$4.1 & 43 & 30.3 & 14.1 & \nodata & \nodata & \nodata & \nodata \\
1237.544 & H2 2-2R11 & \phantom{$-$}0.35 & 95,229 & 0.8 & 31 & 3.9 & 1.8 & \nodata & \nodata & \nodata & \nodata \\
1237.880 & H2 1-2P8 & $-$0.30 & 92,516 & $-$4.0 & 31 & 14.8 & 6.8 & \nodata & \nodata & \nodata & \nodata \\
1241.890 & S I & \nodata & 80,521 & 9.2 & 38 & 4.8 & 2.2 & \nodata & \nodata & \nodata & \nodata \\
1247.160 & S I & $-$0.79 & 80,182 & $-$2.5 & 32 & 7.7 & 3.5 & \nodata & \nodata & \nodata & \nodata \\
1253.325 & S I & $-$1.06 & 80,184 & $-$2.0 & 28 & 6.1 & 2.8 & \nodata & \nodata & \nodata & \nodata \\ 
1253.659 & H2 4-4 R3 & $-$0.45 & 95,584 & $-$5.0 & 28 & 4.6 & 2.1 & \nodata & \nodata & \nodata & \nodata \\
1254.129 & H2 3-1 P17 & $-$2.16 & 97,990 & $-$0.7 & 28 & 8.6 & 3.9 & \nodata & \nodata & \nodata & \nodata \\
1256.093 & S I & $-$1.42 & 80,186 & $-$3.0 & 18 & 2.1 & 1.0 & \nodata & \nodata & \nodata & \nodata \\ 
1257.826 & H2 1-3R3 & $-$0.12 & 91,886 & $-$1.2 & 28 & 52.3 & 23.8 & $-$4.3 & 42 & 4.0 & 0.7 \\
\ldots   & \ldots   & \ldots  &  \ldots  & \ldots  &    & \ldots  & \ldots  & \ldots   &    & \ldots & \ldots \\  
\enddata
\tablecomments{Velocities are all relative to the stellar radial velocity of 20.6 km s$^{-1}$ for \gcru\  and 21.9 for \aori.
    Table \ref{tbl-7} is published in its entirety in the machine-readable
    format. A portion is shown here for guidance regarding its form and
    content.}
\end{deluxetable*}

%\clearpage

%LaTex environment for Table 8, version 3/9/18
%**************************************
%\longrotatetable
\begin{deluxetable*}{lcrrrc|rrr|ccc|rrrr}[h]
\tablecolumns{16}
\footnotesize
\tablecaption{Properties of Chromospheric and Wind Lines in the \gcru\ ASTRAL Spectrum. F$_\mathrm{int}$ is in units of 10$^{-15}$ ergs cm$^{-2}$ sec$^{-1}$; F$_\mathrm{surf}$ is in units of ergs cm$^{-2}$ sec$^{-1}$    \label{tbl-8}}
\tablehead{
\colhead{\lm\ (\AA, vac.)} & Spectrum & \colhead{log\,$gf$} & \colhead{$\tau_\mathrm{lte}$} & \colhead{E$_\mathrm{low}$} & \colhead{E$_\mathrm{high}$} &  
\multicolumn{3}{c}{RV (\kms)} & \multicolumn{3}{c}{FWHM (\kms)} & \colhead{ F$_\mathrm{int} $} & \colhead{ F$_\mathrm{surf}$} & \multicolumn{2}{c}{W$_\lambda$ (\AA)} \\
\colhead{ } & \colhead{ } & \colhead{ } & \colhead{ } & \colhead{ } & \colhead{ } & \colhead{em} & \colhead{abs1} & 
\colhead{abs2} & \colhead{em} & \colhead{abs1} & \colhead{abs2} & \colhead{em} & \colhead{em} & \colhead{abs1} &\colhead{abs2} 
}
\startdata 
 1244.535 & C I & $-$3.80 & $-$4.8 & 10,193 & 90,544 & $-$4.1 & \nodata & \nodata & 26 & \nodata & \nodata & 4.4 & 2.0 & \nodata & \nodata \\
 1245.183 & C I & $-$3.90 & $-$4.9 & 10,193 & 90,502 & $-$1.6 & \nodata & \nodata & 19 & \nodata & \nodata & 3.3 & 1.5 & \nodata & \nodata \\
 1245.964 & C I & $-$3.45 & $-$4.4 & 10,193 & 90,452 & 6.7 & \nodata & \nodata & 35 & \nodata & \nodata & 13.3 & 6.1 & \nodata & \nodata \\
 1246.176 & C I & $-$3.73 & $-$4.7 & 10,193 & 90,438 & $-$0.6 & \nodata & \nodata & 17 & \nodata & \nodata & 1.7 & 0.8 & \nodata & \nodata \\
 1246.862 & C I & $-$3.57 & $-$4.5 & 10,193 & 90,394 & 0.2 & \nodata & \nodata & 30 & \nodata & \nodata & 5.5 & 2.5 & \nodata & \nodata \\
 1247.867 & C I & $-$3.57 & $-$4.5 & 10,193 & 90,329 & $-$8.1 & \nodata & \nodata & 30 & \nodata & \nodata & 7.0 & 3.2 & \nodata & \nodata \\
 1248.009 & C I & $-$3.67 & $-$4.6 & 10,193 & 90,320 & 1.3 & \nodata & \nodata & 38 & \nodata & \nodata & 8.3 & 3.8 & \nodata & \nodata \\
 1248.993 & C I & $-$3.67 & $-$4.6 & 10,193 & 90,257 & 0.4 & \nodata & \nodata & 22 & \nodata & \nodata & 2.4 & 1.1 & \nodata & \nodata \\
 1249.405 & C I & $-$3.38 & $-$4.3 & 10,193 & 90,231 & $-$1.9 & \nodata & \nodata & 23 & \nodata & \nodata & 4.2 & 1.9 & \nodata & \nodata \\
 1250.403 & C I & $-$3.58 & $-$4.5 & 10,193 & 90,167 & 2.7 & \nodata & \nodata & 17 & \nodata & \nodata & 2.5 & 1.1 & \nodata & \nodata \\
 1251.176 & C I & $-$3.21 & $-$4.2 & 10,193 & 90,117 & $-$1.1 & \nodata & \nodata & 31 & \nodata & \nodata & 6.5 & 3.0 & \nodata & \nodata \\
 1252.208 & C I & $-$3.28 & $-$4.2 & 10,193 & 90,051 & $-$5.0 & \nodata & \nodata & 28 & \nodata & \nodata & 4.5 & 2.1 & \nodata & \nodata \\
 1253.468 & C I & $-$3.47 & $-$4.4 & 10,193 & 89,971 & $-$3.4 & \nodata & \nodata & 37 & \nodata & \nodata & 6.2 & 2.8 & \nodata & \nodata \\
 1254.525 & C I & $-$3.16 & $-$4.1 & 10,193 & 89,904 & $-$3.3 & \nodata & \nodata & 23 & \nodata & \nodata & 3.3 & 1.5 & \nodata & \nodata \\
 1256.498 & C I & $-$2.96 & $-$3.9 & 10,193 & 89,779 & 4.9 & \nodata & \nodata & 37 & \nodata & \nodata & 7.3 & 3.3 & \nodata & \nodata \\
 1257.578 & C I & $-$2.94 & $-$3.9 & 10,193 & 89,710 & $-$2.5 & \nodata & \nodata & 31 & \nodata & \nodata & 4.0 & 1.8 & \nodata & \nodata \\
 1258.798 & Si I & $-$0.35 & $-$0.3 & 223 & 79,663 & $-$8.1 & \nodata & \nodata & 27 & \nodata & \nodata & 3.5 & 1.6 & \nodata & \nodata \\
 1261.737 & C I & $-$2.95 & $-$3.9 & 10,193 & 89,448 & $-$6.7 & \nodata & \nodata & 33 & \nodata & \nodata & 5.4 & 2.4 & \nodata & \nodata \\
 1267.596 & C I & $-$2.92 & $-$3.9 & 10,193 & 89,082 & 0.6 & \nodata & \nodata & 22 & \nodata & \nodata & 4.8 & 2.2 & \nodata & \nodata \\
 1274.756 & C I & $-$3.33 & $-$4.3 & 10,193 & 88,639 & $-$2.0 & \nodata & \nodata & 23 & \nodata & \nodata & 9.6 & 4.3 & \nodata & \nodata \\
 1276.287 & C I & $-$2.73 & $-$3.7 & 10,193 & 88,545 & $-$1.2 & \nodata & \nodata & 29 & \nodata & \nodata & 9.0 & 4.0 & \nodata & \nodata \\
 1288.037 & C I & $-$3.67 & $-$4.6 & 10,193 & 87,830 & $-$5.1 & \nodata & \nodata & 26 & \nodata & \nodata & 5.4 & 2.4 & \nodata & \nodata \\
 1288.422 & C I & $-$1.54 & $-$2.5 & 10,193 & 87,807 & $-$1.0 & \nodata & \nodata & 31 & \nodata & \nodata & 12.2 & 5.4 & \nodata & \nodata \\
 1289.984 & C I & $-$2.44 & $-$3.4 & 10,193 & 87,713 & $-$4.4 & \nodata & \nodata & 28 & \nodata & \nodata & 8.4 & 3.7 & \nodata & \nodata \\
 1291.304 & C I & $-$2.32 & $-$3.3 & 10,193 & 87,634 & 1.2 & \nodata & \nodata & 23 & \nodata & \nodata & 4.4 & 2.0 & \nodata & \nodata \\
 1302.168 & O I & $-$0.59 & $-$0.5 & 0 & 76,795 & $-$2.2 & $-$4.2 & $-$31.2 & 64 & 54 & 17 & 8320.3 & 3681.7 & 150.4 & 3.4 \\
 1304.858 & O I & $-$0.81 & $-$0.7 & 158 & 76,794 & $-$0.1 & 5.2 & $-$16.1 & 50 & 31 & 27 & 7447.4 & 3290.7 & 10.9 & 6.1 \\
 1306.029 & O I & $-$1.29 & $-$1.2 & 227 & 76,794 & $-$2.4 & 8.5 & $-$11.9 & 54 & 24 & 33 & 5984.1 & 2642.5 & 35.3 & 86.0 \\
 \ldots   &     & \ldots  & \ldots &     & \ldots & \ldots & \ldots & \ldots & &    &    & \ldots & \ldots &      &      \\
\enddata
\tablecomments{Table \ref{tbl-8} is published in its entirety in the
    machine-readable format. A portion is shown here for guidance regarding
    its form and content.}
\end{deluxetable*}
 %************************************  

%\clearpage

%LaTex environment for Table 9, version 3/9/18      
%\longrotatetable
\begin{deluxetable*}{lcrrrc|rrr|ccc|rrrr}[hb]
\tablecolumns{16}
\movetabledown=1.25in
\footnotesize
\tablecaption{Properties of Chromospheric and Wind Lines in the \aori\ ASTRAL Spectrum. F$_\mathrm{int}$ is in units of 10$^{-15}$ ergs cm$^{-2}$ sec$^{-1}$; F$_\mathrm{surf}$ is in units of ergs cm$^{-2}$ sec$^{-1}$   \label{tbl-9}}
\tablehead{
\colhead{\lm\ (\AA, vac.)} & Spectrum & \colhead{log\,$gf$} & \colhead{$\tau_\mathrm{lte}$} & \colhead{E$_\mathrm{low}$} & \colhead{E$_\mathrm{high}$} &  
\multicolumn{3}{c}{RV (\kms)} & \multicolumn{3}{c}{FWHM (\kms)} & \colhead{ F$_\mathrm{int} $} & \colhead{ F$_\mathrm{surf}$} & \multicolumn{2}{c}{W$_\lambda$ (\AA)} \\
\colhead{ } & \colhead{ } & \colhead{ } & \colhead{ } & \colhead{ } & \colhead{ } & \colhead{em} & \colhead{abs1} & 
\colhead{abs2} & \colhead{em} & \colhead{abs1} & \colhead{abs2} & \colhead{em} & \colhead{em} & \colhead{abs1} & \colhead{abs2} 
       }
\startdata 
 1245.964 & C I & $-$3.45 & $-$4.4 & 10,193 & 90,452 & $-$3.6 & \nodata & \nodata & 15 & \nodata & \nodata & 0.3 & 0.1 & \nodata & \nodata \\
 1246.176 & C I & $-$3.73 & $-$4.7 & 10,193 & 90,438 & $-$31.8 & \nodata & \nodata & 59 & \nodata & \nodata & 2.2 & 0.4 & \nodata & \nodata \\
 1302.168 & O I & $-$0.59 & $-$0.5 & 0 & 76,795 & \nodata & \nodata & \nodata & \nodata & \nodata & \nodata & \nodata & \nodata & \nodata & \nodata \\
 1302.862 & S I & $-$1.07 & $-$1.0 & 396 & 77,150 & $-$1.7 & $-$3.7 & 12.4 & 41 & 29 & 18  & 109.0 &  19.2  & 30.1 & 4.6 \\
 1304.858 & O I & $-$0.84 & $-$0.7 & 158 & 76,794 & 10.9   & 0.7     & 25.4   & 65 & 48 & 32 & 157.8  & 27.8   & 10.9 & 3.2 \\
 1306.029 & O I & $-$1.32 & $-$1.2 & 227 & 76,794 & $-$3.4 & $-$24.4 &  7.6   & 72 & 44 & 44 & 253.5  & 44.5   & 8.4  & 10.5 \\
 \ldots   &     & \ldots  & \ldots &     & \ldots & \ldots & \ldots  & \ldots &    &    &    & \ldots & \ldots &      &      \\
\enddata
\tablecomments{Table \ref{tbl-9} is published in its entirety in the
    machine-readable format. A portion is shown here for guidance regarding
    its form and content.}
\end{deluxetable*}
%************************************   

\end{document}